\documentclass[11pt]{article}

\usepackage{color}
\usepackage{amsmath,amssymb}
\usepackage{graphicx}
\usepackage{graphics}
\usepackage{dcolumn}
\usepackage{bm}
\usepackage{afterpage}

\usepackage[a4paper,left=1.75cm,right=1.75cm,top=3.0cm,bottom=3.0cm]{geometry}

\numberwithin{equation}{section}

\begin{document}

\allowdisplaybreaks

\title{Black Holes and Boson Stars with One Killing Field in Arbitrary Odd Dimensions}

\vskip1cm
\author{Sean Stotyn, Miok Park, Paul L. McGrath and Robert B. Mann\\ \\ \it{Department of Physics and Astronomy, University of Waterloo,}\\ \it{Waterloo, Ontario, Canada, N2L 3G1}\\ \\
                   \small{smastoty@sciborg.uwaterloo.ca, m7park@sciborg.uwaterloo.ca,}\\ \small{pmcgrath@sciborg.uwaterloo.ca, rbmann@sciborg.uwaterloo.ca}}

\date{}

\maketitle

\begin{abstract}

We extend the recent $D=5$ results of Dias, Horowitz and Santos by finding asymptotically AdS rotating black hole and boson star solutions with scalar hair in arbitrary odd spacetime dimension.  Both the black holes and the boson stars are invariant under a single Killing vector field which co-rotates with the scalar field and, in the black hole case, is tangent to the generator of the horizon.  Furthermore, we explicitly construct boson star and small black hole ($r_+ \ll \ell$) solutions perturbatively assuming a small amplitude for the scalar field, resulting in solutions valid for low energies and angular momenta.  We find that just as in $D=5$, the angular momentum is primarily carried by the scalar field in $D>5$, whereas unlike $D=5$ the energy is also primarily carried by the scalar field in $D>5$; the thermodynamics in $D=5$ are governed by both the black hole and scalar field whereas in $D>5$ they are governed primarily by the scalar field alone.  We focus on cataloguing these solutions for the spacetime dimensions of interest in string theory, namely $D=5,7,9,11$.  

\end{abstract}

\newpage

\section{Introduction}

In general, finding analytic solutions to Einstein's equations is not an easy task; a common strategy is to assume a set of symmetries and input a suitable ansatz into the field equations.  This is not guaranteed \emph{a priori} to produce a consistent solution but there exist theorems on spacetime structure which aid in choosing an appropriate ansatz.  For instance, it is well-understood that in any number of dimensions a stationary spacetime must also be axisymmetric \cite{Hawking:1971vc,Hollands:2006rj,Moncrief:2008mr}, that four-dimensional black holes can only have a spherical topology \cite{Friedman:1993ty}, or that black holes in spacetimes with non-negative cosmological constant have no hair \cite{Ruffini:1971xx,Bhattacharya:2007ap} when coupled to ``ordinary'' Maxwell-type matter fields.

Static black holes without rotational symmetry  were constructed perturbatively in situations where the horizon radius is close to the critical radius for  instability of the Reissner-Nordstr\"om solution \cite{Ridgway:1995ke}.
 Essentially 
a magnetically charged Reissner-Nordstr\"om black hole whose 
 horizon radius is less than a critical value near its inverse mass is unstable classically  against the development of a nonzero vector meson field just outside the horizon.  Such black holes have   vector meson hair. 
For spacetimes with a negative cosmological constant the no-hair theorem no longer applies because the reflecting boundary conditions of AdS can support a non-trivial matter field.  Despite this, all known asymptotically AdS black hole solutions had at least two Killing vectors, regardless of the presence of hair. 
Recently a novel solution was constructed \cite{Dias:2011at}  that   considered five-dimensional Einstein gravity  with negative cosmological constant minimally coupled to a massless complex doublet scalar field, describing ``lumpy'' scalar hair co-rotating with a black hole.  This configuration is motivated by superradiance: scalar fields in AdS spaces can increase their amplitude by scattering off the horizon of a rotating black hole, which is then reflected back to the horizon by the AdS boundary conditions, leading to a further increase in amplitude.  This process mines rotational energy from the black hole and the end result is lumpy scalar hair co-rotating at the same angular velocity with the black hole.  The spacetime and scalar fields are collectively invariant under a single Killing vector, which is tangent to the generator of the horizon.

In this paper, we extend the results of \cite{Dias:2011at} to arbitrary odd spacetime dimension $D\ge5$ and construct analytic boson star and black hole solutions perturbatively in the dimensionless scalar amplitude parameter $\epsilon\ll1$.  We do so by a cohomogeneity-1 ansatz, i.e. metric functions of the radial coordinate only, and a judicious choice of scalar fields whose stress tensor shares the symmetries of the metric.  This ensures that the resulting equations of motion form a set of coupled ODEs instead of a system of coupled PDEs.  We catalogue the results for the spacetime dimensions of interest in string theory, $D=5,7,9,11$.  We include the $D=5$ results in the interest of having our paper be self-contained and because we find various discrepancies between our results and the results of \cite{Dias:2011at}.  These discrepancies are explicitly discussed when they appear in {\S}\ref{NearRegion} and {\S}\ref{MatchingConditions}.

The rest of the paper is outlined as follows.  In {\S}\ref{Setup} we introduce the metric and scalar field ansatz and give the set of ODEs following from the equations of motion.  Next, in {\S}\ref{Boson}, we construct perturbative solutions for the boson star in AdS using the scalar field amplitude as an expansion parameter.  Perturbative small black hole solutions follow in {\S}\ref{BlackHole} by introducing a second expansion parameter, $r_+ / \ell \ll 1$.  It is interesting to note that this two-parameter class of black hole solutions is related to both the boson star ($r_+ \rightarrow 0$) and Myers-Perry ($\epsilon \rightarrow 0$) AdS solutions \cite{Myers:1986un}.  The thermodynamic properties of both the boson star and black hole solutions are discussed in {\S}\ref{Thermo}.  Finally, we conclude in {\S}\ref{Discussion} with a discussion of our results and of future work needed in this area.

\section{Setup}\label{Setup}

In this section we introduce the model for constructing hairy black holes and boson stars in arbitrary odd dimension and give the resulting equations of motion.

\subsection{Metric and Scalar Field Ansatz}

We begin with $D=n+2$ dimensional Einstein gravity with negative cosmological constant minimally coupled to an $\frac{n+1}{2}$-tuplet complex scalar field
\begin{equation}
S=\frac{1}{16\pi}\int{d^Dx\sqrt{-g}\left(R+\frac{n(n+1)}{\ell^2}-2\big|\nabla\vec{\Pi}\big|^2\right)}\label{eq:action}
\end{equation}
where we take the usual convention $\Lambda=-\frac{n(n+1)}{2\ell^2}$.  In order to obtain the desired symmetries in our solution, namely that the matter stress tensor has the same symmetries as the metric, we will need to Hopf fibrate our $n$-sphere.  Thus, we consider only odd dimensions with $n\ge3$ and propose the metric and scalar field ansatz
\begin{equation}
ds^2=-f(r)g(r)dt^2+\frac{dr^2}{f(r)}+r^2\bigg(h(r)\big(d\chi+A_idx^i-\Omega(r) dt\big)^2+g_{ij}dx^idx^j\bigg)\label{eq:metric}
\end{equation}
\begin{equation}
\Pi_i=\Pi(r) e^{-i\omega t}z_i, \quad\quad\quad\quad i=1...\frac{n+1}{2} \label{eq:ScalarField}
\end{equation}
where $z_i$ are complex coordinates such that $\displaystyle\sum_{i}dz_id\bar{z}_i$ is the metric of a unit $n-$sphere.  An explicit and convenient choice for the $z_i$ is
\begin{equation}
z_i=\left\{ {\genfrac{.}{.}{0pt}{0}{e^{i(\chi+\phi_i)}\cos\theta_i\displaystyle\prod_{j<i}\sin\theta_j,\quad\quad\quad i=1...\frac{n-1}{2}}{{e^{i\chi}\displaystyle\prod_{j=1}^{\frac{n-1}{2}}\sin\theta_j},\quad\quad\quad\quad\quad i=\frac{n+1}{2}}}\right. \label{eq:zi}
\end{equation}
in which case $\displaystyle\sum_{i}dz_id\bar{z}_i=(d\chi+A_idx^i)^2+g_{ij}dx^idx^j$ is the Hopf fibration of the unit $n-$sphere where
\begin{equation}
A_idx^i=\sum_{i=1}^{\frac{n-1}2}{\cos^2\theta_i\left[\prod_{j<i}\sin^2\theta_j\right]d\phi_i}
\end{equation}
and $g_{ij}$ is the metric on a unit $\mathbb{CP}^{\frac{n-1}{2}}$.  In these coordinates the scalar fields are manifestly single-valued on the spacetime since $\chi$ and $\phi_i$ have period $2\pi$ while the $\theta_i$ have period $\frac{\pi}{2}$.  To verify our ansatz at this point, we note that if we choose $n=3$ and perform the coordinate transformation $\chi=\psi-\frac{\phi}{2},$ $\theta=\frac{\vartheta}2$ we recover exactly the ansatz considered in Ref. \cite{Dias:2011at}.

The form of the scalar fields is crucial to this construction and was first considered in \cite{Hartmann:2010pm}: it is clear from Eq. (\ref{eq:ScalarField}) that the scalar fields can be viewed as coordinates on ${\mathbb C}^\frac{n+1}{2}$.  Given that $\Pi(r)$ is a function of $r$ only, for each value of $r$, $\vec{\Pi}$ traces out a round $n$-sphere with a  time-varying but otherwise constant  phase.  On the other hand, constant $r$ surfaces in the metric (\ref{eq:metric}) correspond to squashed rotating $n$-spheres.  The stress tensor for the scalar field takes the form
\begin{equation} \label{eq:Tab}
T_{ab}=\left(\partial_a\vec{\Pi}^*\partial_b\vec{\Pi}+\partial_a\vec{\Pi}\partial_b\vec{\Pi}^*\right)-g_{ab}\left(\partial_c\vec{\Pi}\partial^c\vec{\Pi}^*\right)
\end{equation}
which has the same symmetries as the metric (\ref{eq:metric}) since the first term is the pull-back of the round metric of the $n$-sphere and the second term is proportional to $g_{ab}$.

Although the matter stress tensor has the same symmetries as the metric, the scalar fields themselves do not.  Indeed, the metric (\ref{eq:metric}) is invariant under $\partial_t$, $\partial_\chi$ as well as the rotations of $\mathbb{CP}^{\frac{n-1}{2}}$ while the scalar field (\ref{eq:ScalarField}) is only invariant under the combination
\begin{equation}
K=\partial_t+\omega\partial_\chi. \label{eq:KV}
\end{equation}
Therefore, any solution with non-trivial scalar field will only be invariant under the single Killing vector field given by (\ref{eq:KV}).

\subsection{Equations of Motion}

The equations of motion resulting from the action (\ref{eq:action}) are $G_{ab}-\frac{n(n+1)}{2\ell^2}g_{ab}=T_{ab}$ and $\nabla^2\vec{\Pi}=0$ which ought to have non-trivial solutions by virtue of the matter stress tensor possessing the same symmetries as the metric.  Indeed, inserting the ansatz (\ref{eq:metric}) and (\ref{eq:ScalarField}) into the equations of motion yields a system of five coupled second order ODEs
\begin{equation}
\begin{split}
f''-\frac{6f'}{fr}\left(\frac{rf'}{6}-\frac f6+\Xi\right) + \frac{4h'}{r}+\frac{n^2-1}{r^2}+\frac{n^2-1}{\ell^2}+\frac{8\Pi'\Pi}{r}+\frac{4\Pi^2(\omega-\Omega)^2}{fg}&\\-\frac{8\Xi^2}{fr^2}
-\frac{4\Pi^2\left(1+\frac{(n-1)}2h\right)}{hr^2}-\frac{2(n-3)\Xi}{r^2}=0&\label{eq:fEq}
\end{split}
\end{equation}
\begin{equation}
\begin{split}
g''-g'\left(\frac{4\Xi}{fr}+\frac{g'}{g}-\frac{1}{r}\right)-4g\left(\frac{\big(\Xi r^{\frac{n-1}2}\sqrt{h}\big)'}{fr^{\frac{n+1}2}\sqrt{h}}+\frac{\frac{(n-1)}2h^2-\Pi^2}{fhr^2}+\frac{3(n+1)}{2f\ell^2}-\frac{(n-3)\Xi}{2fr^2}\right)&\\
-\frac{8\Pi^2(\omega-\Omega)^2}{f^2}-\frac{hr^2\Omega'^2}{f}=0&\label{eq:gEq}
\end{split}
\end{equation}
\begin{equation}
h''+\frac{h'}{r}-\frac{2h'}{fr}\left(\Xi+\frac{frh'}{2h}\right)+\frac{h^2r^2\Omega'^2}{fg}+\frac{4(1-h)}{fr^2}\left(\Pi^2+\frac{(n+1)}{2}h\right)=0\label{eq:hEq}
\end{equation}
\begin{equation}
\Omega''+\frac{4\Pi^2}{fhr^2}(\omega-\Omega)+\Omega'\left(\frac{f'}{f}+\frac{2h'}{h}+\frac{2\Xi}{fr}+\frac{2n+1}{r}\right)=0\label{eq:OmegaEq}
\end{equation}
\begin{equation}
\Pi''-\frac{2\Pi'}{fr}\left(\Xi-\frac{f}{2}\right)+\frac{\Pi(\omega-\Omega)^2}{f^2g}-\frac{\big(1+(n-1)h\big)\Pi}{fhr^2}=0\label{eq:PiEq}
\end{equation}
where $\Xi=h+\Pi^2-\frac{n+1}{2}-\frac{(n+1)r^2}{2\ell^2}$ and a $'$ denotes differentiation with respect to $r$.  In addition to these second order ODEs, the Einstein equations further impose two first order ODEs in the form of constraint equations, $C_1=0$ and $C_2=0$. Explicitly, these are
\begin{equation}
C_1=\frac{(f^2ghr^{2(n-1)})'}{fr^{2n-3}}+4gh\Xi
\end{equation}
\begin{equation}
C_2=\frac{\Pi^2(\omega-\Omega)^2}{f^2g}+\Pi'^2-\frac{r^2h\Omega'^2}{4fg}+\frac{\Xi(r^{n+1}h)'}{fhr^{n+2}}+\frac{(hf)'h'}{4fh^2}+\frac{n(fhr^{n-1})'}{2fhr^n}+\frac{\frac{(n-1)}{2}h^2-\Pi^2}{fhr^2}+\frac{n+1}{2f\ell^2}.
\end{equation}

We note that inserting $n=3$ into the above equations of motion yields the 5D results of \cite{Dias:2011at}.  Additionally, it is interesting to note the presence of terms proportional to $n-3$ in Eqs. (\ref{eq:fEq}) and (\ref{eq:gEq}), which are therefore absent in 5 dimensions; one might expect such terms to change the physics of the higher dimensional solutions,
though at the perturbative level this may not be apparent.
  Before we move on, we emphasize  that the above equations of motion are \emph{exact} for arbitrary odd dimension $D\ge5$; we have explicitly verified this for $D=5,7,9,11$.

\section{Perturbative Boson Stars}\label{Boson}

In this section, we  present the boundary conditions that define a boson star and then use these to  construct such solutions as perturbations around AdS.  The boson star is a horizonless solution for the matter configuration we are using and, furthermore, can be viewed as a warm-up problem to the hairy black hole solutions we will later construct.  The expansion  is carried out in orders of the scalar field condensate parameter $\epsilon$ and we give results up to order $\epsilon^6$ for the spacetime dimensions of interest in string theory, namely $D=5,7,9,11$.  As a perturbative construction, these results will only be valid for small energies and angular momenta.

\subsection{Boson Star Boundary Conditions} \label{BosonConditions}

\subsubsection*{Boson Star Origin}

Boson stars are smooth, horizonless geometries, which means that all metric functions must be regular at the origin.  Furthermore, due to the slow physical rotation of points as $r\rightarrow0$, surfaces of constant $t$ in the vicinity of the origin ought to be described by round $n$-spheres with $r$ being the proper radial distance.  To find the boundary condition on $\Pi$, we multiply (\ref{eq:PiEq}) by $r^2$ and note that $\Pi$ must vanish at the origin in order to yield consistent equations of motion.  Thus, the boundary conditions at the boson star origin take the form
\begin{eqnarray}
\left.f\right|_{r \rightarrow 0} = 1 + \mathcal{O}(r^2), \quad 
\left.g\right|_{r \rightarrow 0} = g(0)+\mathcal{O}(r), \quad 
\left.h\right|_{r \rightarrow 0} = 1 + \mathcal{O}(r^2), \label{eq:OriginBC}\\
\left.\Omega\right|_{r \rightarrow 0} = \Omega(0)+\mathcal{O}(r), \quad 
\left.\Pi\right|_{r \rightarrow 0} = \mathcal{O}(r).\quad\quad\quad\quad\quad \nonumber
\end{eqnarray}

\subsubsection*{Boson Star Asymptotics}

In order to simplify the asymptotic boundary conditions, we first make note of a residual gauge freedom.  It is straightforward to show that the transformation
\begin{eqnarray} \label{PsiGaugeFreedom}
\chi \rightarrow \chi + \lambda t, \quad\quad \Omega \rightarrow \Omega + \lambda, \quad\quad \omega \rightarrow \omega - \lambda
\end{eqnarray}
for some arbitrary constant $\lambda$, leaves both the metric (\ref{eq:metric}) and scalar field (\ref{eq:ScalarField}) unchanged.  We will use this gauge invariance to pick a frame which is not rotating at infinity, i.e. we use it to set $\Omega \rightarrow 0$ in the limit $r \rightarrow \infty$.

In the $r \rightarrow \infty$ limit the boundary conditions for the boson star will asymptote to AdS with corrections for mass and angular momentum.  To ensure the solution has a Newtonian potential of the correct strength for a spherically symmetric mass distribution, we impose an $r^{-(n-1)}$ correction for $f$.  Next, requiring our solutions to have finite masses and angular momenta means we must impose an $r^{-(n+1)}$ fall-off for the corrections to $g,h$ and $\Omega$.  These considerations determine the boundary conditons for $f,g,h,$ and $\Omega$ up to some constants $C_f, C_h,$ and $C_\Omega$.  The remaining boundary condition is set by requiring $\Pi$ to be normalizable, which means it must decay like $r^{-(n+1)}$.  Explicitly, the asymptotic boundary conditions are given by
\begin{align} \label{eq:AsymBC}
\left.f\right|_{r \rightarrow \infty}={}& \frac{r^2}{\ell^2} + 1 + \frac{C_f \ell^{n-1}}{r^{n-1}} + \mathcal{O}(r^{-n}), \quad\>\> 
\left.g\right|_{r \rightarrow \infty} = 1 - \frac{C_h \ell^{n+1}}{r^{n+1}} + \mathcal{O}(r^{-(n+2)}), \nonumber \\ 
\left.h\right|_{r \rightarrow \infty}={}& 1 + \frac{C_h \ell^{n+1}}{r^{n+1}}+ \mathcal{O}(r^{-(n+2)}), \quad\quad 
\left.\Omega\right|_{r \rightarrow \infty} = \frac{C_{\Omega} \ell^{n}}{r^{n+1}} + \mathcal{O}(r^{-(n+2)}),\\ 
\left.\Pi\right|_{r \rightarrow \infty}={}& \frac{\epsilon \ell^{n+1}}{r^{n+1}} + \mathcal{O}(r^{-(n+2)}).\nonumber
\end{align}
Here and in what follows, $\epsilon$ provides a dimensionless measure of the amplitude of the scalar field.

With the appropriate boundary conditions and the equations of motion at hand, we are now ready to construct perturbative boson stars in arbitrary odd dimensions.

\subsection{Perturbative Boson Star}\label{PertBS}

We start by expanding our fields in terms of the scalar field amplitude, $\epsilon$, as follows:
\begin{equation}
F(r,\epsilon)=\sum_{i=0}^m{\tilde{F}_{2i}(r)\epsilon^{2i}}\quad\quad\quad \Pi(r,\epsilon)=\sum_{i=0}^m{\tilde{\Pi}_{2i+1}(r)\epsilon^{2i+1}}\quad\quad\quad \omega(\epsilon)=\sum_{i=0}^m{\tilde{\omega}_{2i}\epsilon^{2i}} \label{eq:FieldExpansion}
\end{equation}
where $F=\{f,g,h,\Omega\}$ is shorthand for each of the metric functions in (\ref{eq:metric}).  The metric functions are expanded in even powers of $\epsilon$ while the scalar fields are expanded in odd powers.  This allows a perturbative expansion as follows: start with global AdS at $m=0$, then introduce non-trivial scalar fields into the AdS background, without back-reacting on the metric, by solving (\ref{eq:PiEq}).  The full set of equations of motion will then be satisfied up to order $\epsilon$.  At next order, $m=1$, the scalar fields due to $\tilde{\Pi}_1(r)$ then source corrections to the gravitational fields $\tilde{F}_2(r)$, and these in turn back-react on the scalar fields $\tilde{\Pi}_3(r)$.  The equations of motion will then be satisfied up to order $\epsilon^3$.  The perturbative solution can, in principle, be obtained by this bootstrapping procedure up to arbitrary order, $m$.  Note that we also must expand the frequency in even powers of $\epsilon$.  This is because at the linear order, the frequency is determined by the scalar field alone but at the next order, the scalar field then back reacts on the metric inducing non-trivial frame-dragging effects which in turn have a non-trivial effect on the rotation of the scalar field.  In practice, these corrections to $\omega$ are found by imposing the boundary conditions.

Global AdS is given by
\begin{equation}
f_0=1+\frac{r^2}{\ell^2},\quad\quad g_0=1,\quad\quad h_0=1,\quad\quad \Omega_0=0
\end{equation}
In this background, the most general massless scalar field solution to (\ref{eq:PiEq}) which is consistent with the asymptotic boundary conditions (\ref{eq:AsymBC}) is given by
\begin{equation}
\Pi_1(r)=\frac{r\ell^{n+1} }{(r^2+\ell^2)^{\frac{n+2}{2}}}{_2F_1}\left[\frac{n+2-\omega \ell}{2},\frac{n+2+\omega\ell}{2};\frac{n+3}{2};\frac{\ell^2}{r^2+\ell^2}\right]\label{eq:Pi1}
\end{equation}
where $_2F_1$ is the hypergeometric function.  Now in order to satisfy the boundary conditions at the origin (\ref{eq:OriginBC}) we must further restrict $\omega$ to
\begin{equation}
\omega\ell=n+2+2k,\quad\quad\quad\quad k=0,1,2,...
\end{equation}
where the non-negative integer, $k$, describes the various possible radial modes of the scalar field.  Although in principle, any radial profile can be built up out of a linear combination of the radial modes, this introduces multiple frequency parameters, $\omega_k$.  This is inconsistent with the existence of the Killing vector field (\ref{eq:KV}).  Furthermore, the mode $k=0$ yields the ground state while higher modes represent excited states.  In what follows, we therefore choose $k=0$ as the only mode present, in which case (\ref{eq:Pi1}) simplifies to
\begin{equation}
\Pi_1(r)=\frac{r\ell^{n+1} }{(r^2+\ell^2)^{\frac{n+2}{2}}}. \label{eq:Piepsilon}
\end{equation}

Proceeding up the perturbative ladder, we insert (\ref{eq:Piepsilon}) and the expansion (\ref{eq:FieldExpansion}) into the equations of motion, expand in $\epsilon$ and solve for order $\epsilon^2$.  In general, the solutions contain two constants of integration, which are then uniquely fixed by the boundary conditions.  These fields, $\tilde{F}_2(r)$ are then inserted into the equation of motion for $\Pi(r)$ to find the $\epsilon^3$ correction to the scalar fields.  This process can, in principle, be taken to arbitrary order in $\epsilon$. In practice, however, the expressions become rather unwieldy at higher orders making this increasingly difficult to accomplish.  Up to order $\epsilon^6$, we find the general solutions to take the form
\begin{equation}
f(r)=1+\frac{r^2}{\ell^2}-\frac{r^2\ell^{n-1}f_{n;2,0}}{(r^2+\ell^2)^{n+1}}\epsilon^2-\frac{r^2\ell^{n-1}f_{n;4,0}}{(r^2+\ell^2)^{2n+3}}\epsilon^4+{\cal O}(\epsilon^6)\label{eq:fExpansion}
\end{equation}
\begin{equation}
g(r)=1-\frac{2\ell^{2n+2}\big((n+1)r^2+\ell^2\big)}{n(r^2+\ell^2)^{n+2}}\epsilon^2-\frac{\ell^{n+1}g_{n;4,0}}{(r^2+\ell^2)^{2n+4}}\epsilon^4+{\cal O}(\epsilon^6)
\end{equation}
\begin{equation}
h(r)=1+\frac{r^2\ell^{n+1}h_{n;4,0}}{(r^2+\ell^2)^{2n+3}}\epsilon^4+{\cal O}(\epsilon^6)
\end{equation}
\begin{equation}
\Omega(r)=\frac{\ell^{n}\Omega_{n;2,0}}{(r^2+\ell^2)^{n+1}}\epsilon^2+\frac{\ell^{n}\Omega_{n;4,0}}{(r^2+\ell^2)^{2n+3}}\epsilon^4+{\cal O}(\epsilon^6)
\end{equation}
\begin{equation}
\Pi(r)=\frac{r\ell^{n+1} }{(r^2+\ell^2)^{\frac{n+2}{2}}}\epsilon+\frac{r\ell^{n+3}\Pi_{n;3,0}}{(r^2+\ell^2)^{\frac{3n+4}{2}}}\epsilon^3+\frac{r\ell^{n+3}\Pi_{n;5,0}}{(r^2+\ell^2)^{\frac{5(n+2)}{2}}}\epsilon^5+{\cal O}(\epsilon^7)\label{eq:PiExpansion}
\end{equation}
where the fields $\{f_{n;s,0},g_{n;s,0},h_{n;s,0},\Omega_{n;s,0},\Pi_{n;s,0}\}$ are simple polynomials in $r$; in this notation $s$ labels the order in $\epsilon$ and $n=D-2$ labels the spacetime dimension\footnote{The purpose of the ``$0$'' index on these coefficients will be clear when we compare to the perturbative black hole solutions which involves a second pertubation parameter.}.  These fields are catalogued in Appendix B for $n=3,5,7,9$ up to order $\epsilon^6$.  Similarly, the explicit corrections to $\omega_n$ for the boson star can be found by taking the $r_+\rightarrow0$ limit of the general expressions for $\omega_n$ in Appendix A.

\section{Perturbative Black Holes}\label{BlackHole}

We now turn our attention to the more interesting hairy black hole solutions.  The presence of a horizon at $r=r_+$ provides us with an additional length scale, or rather, another perturbative expansion parameter, $r_+ / \ell$, on top of the scalar field amplitude, $\epsilon$, making this a two-parameter family of solutions.  To find the black hole solution then, we will perform a double expansion in the scalar field condensate parameter $\epsilon$ and dimensionless horizon radius $r_+/\ell$.  As pointed out in \cite{Dias:2011at}, inserting this double expansion into the equations of motion results in a set of differential equations that cannot be simultaneously solved everywhere in the spacetime.  This problem is circumvented by applying a matched asymptotic expansion; this procedure will be made explicit in {\S}\ref{MatchingConditions}.  Basically, this involves splitting the spacetime into two regions, a far-region ($r \gg r_+$) and a near-region ($r_+ \le r \ll \ell$), and solving the equations of motion in each separately.  At each order in our solution we will have two arbitrary constants of integration.  The boundary conditions can be used to fix one of these for each region, i.e. we can apply the asymptotic boundary condition in the far-region and the horizon boundary condition in the near-region.  The remaining constant can be fixed by matching the solutions where the two regions overlap: $r_+ \ll r \ll \ell$.   This admits a unique solution valid in the entire spacetime that satisfies both boundary conditions.   This analysis is carried out in the following section and we give results up to $\mathcal{O}(\epsilon^m (r_{+}/\ell)^n)$ where $m+n \le 6$ for the spacetime dimensions $D=5,7,9,11$.  As in the boson star case we will consider only the ground state hairy black holes with regard to the excitation of the scalar field and, again, these results will only be valid for small energies and angular momenta.

The double expansion of our fields can be interpreted as placing a black hole inside a rotating boson star, or alternatively as placing a non-trivial scalar field around a small rotating black hole.  Consequently, in the limit $r_+\rightarrow0$ we should recover the boson star of the previous section and similarly, in the limit $\epsilon\rightarrow0$ we should recover an asymptotically AdS rotating black hole, i.e. a Myers-Perry-AdS black hole \cite{Myers:1986un,Hawking:1998kw,Gibbons:2004js,Gibbons:2004uw}.  We then see that the two-parameter family of hairy black hole solutions is connected to the two-parameter family of Myers-Perry-AdS solutions.  For the hairy black hole, the frequency is uniquely determined in terms of $r_+$ and $\epsilon$, so in the $\epsilon\rightarrow0$ limit, the hairy black hole joins with a one-parameter subset of Myers-Perry-AdS black holes, whose horizon angular velocity is identified with the $\epsilon\rightarrow0$ limit of $\omega$. To phrase this in terms of the space of solutions, the one-parameter family of boson stars corresponds to a line in the $(\epsilon,\omega)$-plane, the Myers-Perry-AdS black holes correspond to the $(\omega,r_+)$-plane, while the hairy black holes correspond to a sheet through the $(\epsilon,\omega,r_+)$ octant.  This sheet meets the $(\epsilon,\omega)$ plane on the line defining boson stars and it intersects the $(\omega,r_+)$ plane, which is where it joins with the Myers-Perry-AdS family of solutions.  This is represented schematically in the following Table.
\begin{center}
\renewcommand{\arraystretch}{1.5}
\begin{tabular}{ | c |  c | c | c | c| c | }
\hline
Order   &  $\epsilon^0$   &   \quad \quad $\epsilon^2$ \quad \quad & \quad \quad $\epsilon^4$ \quad \quad & $\cdots$ &   $\sum_{i=0}^{\infty} \left(\epsilon\right)^{2i}$  \\ \hline
$\left(\frac{r_+}{\ell}\right)^0$  &      Global AdS  		&    \multicolumn{3}{c|}{$\rightarrow$ Perturbative Boson Star $\rightarrow$} &    Exact Boson Star    \\ \hline
$\left(\frac{r_+}{\ell}\right)^2$  &      $\downarrow$  &  \multicolumn{3}{c|}{\quad $\searrow$ \quad \bf{Perturbative} \quad $\searrow$ \quad} &       \\ \cline{1-1}
$\left(\frac{r_+}{\ell}\right)^4$  &      Pert. MP-AdS BH  &   \multicolumn{3}{c|}{\bf{Hairy}} & $\vdots$      \\ \cline{1-1}
$\vdots$                           &      $\downarrow$    &   \multicolumn{3}{c|}{\quad $\searrow$ \quad \bf{Black Hole} \quad $\searrow$ \quad} &       \\ \hline
$\sum_{i=0}^{\infty} \left(\frac{r_+}{\ell}\right)^{2i}$  &     Exact MP-AdS BH       &   \multicolumn{3}{c|}{$\cdots$}  &   Exact Hairy BH    \\ \hline
\end{tabular}
\end{center}

Before continuing, we explicitly present the Myers-Perry-AdS solution with equal angular momenta in the two planes of rotation.  In terms of the metric ansatz (\ref{eq:metric}), the metric functions take the form \cite{Dias:2011at}
\begin{equation}
f=\frac{r^2}{\ell^2}+1-\frac{r_M^{n-1}}{r^{n-1}}\left(1-\frac{a^2}{\ell^2}\right)+\frac{r_M^{n-1}a^2}{r^{n+1}}, \quad\quad h=1+\frac{r_M^{n-1}a^2}{r^{n+1}}, \quad\quad g=\frac1{h(r)},\quad\quad \Omega=\frac{r_M^{n-1}a}{r^{n+1}h} \label{eq:MPAdS}
\end{equation}
where $r_M$ and $a$ are related to the outer horizon, $r_+$, and the angular velocity of the horizon, $\Omega_H=\omega$, as
\begin{equation}
r_M^{n-1}=\frac{r_+^{n+1}(r_+^2+\ell^2)}{r_+^2\ell^2-a^2(r_+^2+\ell^2)}, \quad\quad\quad a=\frac{r_+^2\ell^2\omega}{r_+^2+\ell^2}.
\end{equation}
Small black holes are then obtained by Taylor expanding the metric functions (\ref{eq:MPAdS}) in powers of $r_+/\ell$.

\subsection{Black Hole Boundary Conditions}

\subsubsection*{Black Hole Horizon}

We wish to find non-extremal black hole solutions with scalar hair, so we begin by defining the location of the non-degenerate outermost horizon to be $r=r_{+}$.  Consequently, $f$ must have a simple zero at $r_+$ while all other metric functions must remain finite and non-zero.  For the scalar field, one might expect $\Pi$ to vanish in the vicinity of the horizon since it was shown in \cite{Pena:1997cy,Astefanesei:2003qy} that one cannot have black holes inside boson stars.  However this applies to static black holes and it was pointed out in \cite{Dias:2011at} that this prohibition is removed if the black hole and the boson star are co-rotating.  Indeed, if we require $\Pi(r_+)$ to be finite and non-zero, then multiplying Eq. (\ref{eq:PiEq}) by $f^2$ shows that the equations of motion remain consistent across the horizon provided $\Omega(r_+)=\omega$.  The boundary conditions at the black hole horizon are thus
\begin{equation}\label{eq:BHBC}
\begin{split}
f\big|_{r \rightarrow r_{+}} = \mathcal{O}(r-r_{+}), &\quad \quad\quad
g\big|_{r \rightarrow r_{+}} = g(r_+)+\mathcal{O}(r-r_+), \quad \quad\quad
h\big|_{r \rightarrow r_{+}} = h(r_+)+\mathcal{O}(r-r_+),\\
&\Omega\big|_{r \rightarrow r_{+}} = \omega+\mathcal{O}(r-r_+), \quad \quad
\Pi\big|_{r \rightarrow r_{+}} =\mathcal{O}(r-r_+).
\end{split}
\end{equation}

\subsubsection*{Black Hole Asymptotics}

The asymptotic boundary conditions for the black hole will be identical to those of the boson star since both are globally AdS with next-to-leading order terms accounting for mass and angular momentum.  As in the case of the boson star, we are also free to exploit the gauge freedom in Eq. (\ref{PsiGaugeFreedom}) and choose to work in a frame which is non-rotating at infinity.  Thus we will apply to the black hole the same asymptotic boundary conditions that we saw for the boson star in Eq. (\ref{eq:AsymBC}).

\subsection{Perturbative Hairy Black Hole}

\subsubsection{Far Region}

As discussed above,   we start by performing a double expansion of our fields  in $\epsilon$ and $r_+ / \ell$ as follows:
\begin{equation}
F^{out}(r,\epsilon,r_+) = \sum_{i=0}^m \sum_{j=0}^n \tilde{F}^{out}_{2j, 2i}(r) \epsilon^{2j} \left(\frac{r_+}{\ell}\right)^{2i}, \quad\quad\quad
\Pi^{out}(r,\epsilon,r_+) = \sum_{i=0}^m \sum_{j=0}^n \tilde{\Pi}^{out}_{2j+1, 2i}(r) \epsilon^{2j+1} \left(\frac{r_+}{\ell}\right)^{2i} \label{eq:FarFieldExpansion}
\end{equation}
where $F^{out}=\{f^{out},g^{out},h^{out},\Omega^{out}\}$ is shorthand for each of the metric functions in the far-region (\ref{eq:metric}).  As discussed in {\S}\ref{PertBS}, the scalar frequency $\omega$ has a similar expansion
\begin{equation}
\omega(\epsilon,r_+) = \sum_{i=0}^m \sum_{j=0}^n \tilde{\omega}_{2j, 2i} \epsilon^{2j} \left(\frac{r_+}{\ell}\right)^{2i} \label{eq:BHomegaExpansion}
\end{equation}
where this expansion holds in both the outer and inner regions since $\omega$ is a globally defined constant. 

The perturbative expansion then proceeds similar to the boson star case except that instead of our background being global AdS we take it to be the Myers-Perry AdS black hole (\ref{eq:MPAdS}) expanded in powers of $r_+ / \ell\ll1$.  We can therefore immediately write down the field coefficients $\tilde{F}^{out}_{0, 2i}(r)$ up to arbitrary order while trivially satisfying the field equations and asymptotic boundary condition.  With the complete $r_+ / \ell$ expansion in hand, the goal then is to introduce a non-trivial scalar field and  solve the field equations  in powers of $\epsilon$.  As before, this is done by first inserting the fields $\tilde{F}^{out}_{0, 2i}(r)$ into Eq. (\ref{eq:PiEq}) and solving the order $\epsilon$ equation to determine $\tilde{\Pi}^{out}_{1,2i}$ order by order in $r_+/\ell$.  The presence of the scalar field then sources the gravitational fields at order $\epsilon^2$, which are then determined by solving Eqs. (\ref{eq:fEq})--(\ref{eq:OmegaEq}) for $\tilde{F}^{out}_{2, 2i}(r)$ order by order in $r_+/\ell$.  At the next order in $\epsilon$, there is a back-reaction on the scalar field, which is then similarly solved for from  Eq. (\ref{eq:PiEq}) order by order in $r_+/\ell$.  This iteration process continues with gravitational field corrections at every even order in $\epsilon$ and scalar field corrections at every odd order in $\epsilon$.  Due to the powers of $r_+/\ell$ that appear in the Myers-Perry solution, we carry out this procedure up to $\mathcal{O}(\epsilon^0 (r_{+}/\ell)^{n+3})$, $\mathcal{O}(\epsilon^2 (r_{+}/\ell)^{n+1})$ and $\mathcal{O}(\epsilon^4 (r_{+}/\ell)^{n-1})$.  Applying the asymptotic boundary conditions we find the fields to have the structure:
\begin{align}
f^{out}(r)={}&\left[\frac{r^2}{\ell^2}+1-\frac{\ell^{n-1}}{r^{n-1}}\frac{r_+^{n-1}}{\ell^{n-1}}-\frac{\big((n+2)^2+1\big)\ell^{n-1}}{r^{n-1}}\frac{r_+^{n+1}}{\ell^{n+1}}+{\cal O}\left(\frac{r_+^{n+3}}{\ell^{n+3}}\right)\right]+\epsilon^2\left[-\frac{r^2\ell^{n-1}f_{n;2,0}}{(r^2+\ell^2)^{n+1}}\right. \label{eq:fFarRegion} \\
&\left.+\frac{\ell^{n-1}f_{n;2,2}^{out}}{r^{n-3}(r^2+\ell^2)^{n+2}}\frac{r_+^{n-1}}{\ell^{n-1}}+{\cal O}\left(\frac{r_+^{n+1}}{\ell^{n+1}}\right)\right]+\epsilon^4\left[-\frac{r^2\ell^{n-1}f_{n;4,0}}{(r^2+\ell^2)^{2n+3}}+{\cal O}\left(\frac{r_+^{n-1}}{\ell^{n-1}}\right)\right], \nonumber\\
g^{out}(r)={}&\left[1+{\cal O}\left(\frac{r_+^{n+3}}{\ell^{n+3}}\right)\right]+\epsilon^2\left[-\frac{2\ell^{2n+2}\big((n+1)r^2+\ell^2\big)}{n(r^2+\ell^2)^{n+2}}+\frac{g_{n;2,2}^{out}}{r^{n-3}(r^2+\ell^2)^{n+3}}\frac{r_+^{n-1}}{\ell^{n-1}}+{\cal O}\left(\frac{r_+^{n+1}}{\ell^{n+1}}\right)\right] \label{eq:gFarRegion} \\
&+\epsilon^4\left[-\frac{\ell^{n+1}g_{n;4,0}}{(r^2+\ell^2)^{2n+4}}+{\cal O}\left(\frac{r_+^{n-1}}{\ell^{n-1}}\right)\right],\nonumber\\
h^{out}(r)={}&\left[1+{\cal O}\left(\frac{r_+^{n+3}}{\ell^{n+3}}\right)\right]+\epsilon^2\left[{\cal O}\left(\frac{r_+^{n+1}}{\ell^{n+1}}\right)\right]+\epsilon^4\left[\frac{r^2\ell^{n+1}h_{n;4,0}}{(r^2+\ell^2)^{2n+3}}+{\cal O}\left(\frac{r_+^{n-1}}{\ell^{n-1}}\right)\right], \label{eq:hFarRegion}\\
\ell\Omega^{out}(r)={}&\left[\frac{(n+2)\ell^{n+1}}{ r^{n+1}}\frac{r_+^{n+1}}{\ell^{n+1}}+{\cal O}\left(\frac{r_+^{n+3}}{\ell^{n+3}}\right)\right]+\epsilon^2\left[\frac{\ell^{n+1}\Omega_{n;2,0}}{(r^2+\ell^2)^{n+1}}+\frac{\Omega_{n;2,2}^{out}}{r^{n-3}(r^2+\ell^2)^{n+2}}\frac{r_+^{n-1}}{\ell^{n-1}}+{\cal O}\left(\frac{r_+^{n+1}}{\ell^{n+1}}\right)\right] \label{eq:OmegaFarRegion} \\
&+\epsilon^4\left[\frac{\ell^{n+1}\Omega_{n;4,0}}{(r^2+\ell^2)^{2n+3}}+{\cal O}\left(\frac{r_+^{n-1}}{\ell^{n-1}}\right)\right]\nonumber\\
\Pi^{out}(r)={}&\epsilon\left[\frac{r\ell^{n+1} }{(r^2+\ell^2)^{\frac{n+2}{2}}}+\frac{\ell^{n+1}\Pi^{out}_{n;1,n-1}}{r^{n-2}(r^2+\ell^2)^{\frac{n+4}{2}}}\frac{r_+^{n-1}}{\ell^{n-1}}+\frac{\ell^{n+1}\Pi_{n;1,n+1}^{out}}{r^n(r^2+\ell^2)^{\frac{n+4}{2}}}\frac{r_+^{n+1}}{\ell^{n+1}}+{\cal O}\left(\frac{r_+^{n+3}}{\ell^{n+3}}\right)\right]\\
&+\epsilon^3\left[\frac{r\ell^{n+3}\Pi_{n;3,0}}{(r^2+\ell^2)^{\frac{3n+4}{2}}}+\frac{\Pi_{n;3,n-1}^{out}}{r^{n-2}(r^2+\ell^2)^{\frac{3(n+2)}{2}}}\frac{r_+^{n-1}}{\ell^{n-1}}+{\cal O}\left(\frac{r_+^{n+1}}{\ell^{n+1}}\right)\right]+\epsilon^5\left[\frac{r\ell^{n+3}\Pi_{n;5,0}}{(r^2+\ell^2)^{\frac{5(n+1)}{2}}}\right.\nonumber\\
&\left.+{\cal O}\left(\frac{r_+^{n-1}}{\ell^{n-1}}\right)\right]\nonumber
\end{align}
where the fields $F_{n;s,0}$ are the boson star fields of the previous section and $F^{out}_{n;s,t}$ are new fields which enter at non-zero order in $r_+/\ell$.  In solving for these fields, one of the arbitrary constants from each second order ODE is fixed by the asymptotic boundary conditions while the other constant is fixed by the matching condition.  The double expansion of $\omega$, which is also obtained by matching the inner and outer solutions, is catalogued in Appendix A and the fields with the matching condition already imposed are catalogued in Appendix B.  
We postpone a discussion of the matching procedure to {\S}\ref{MatchingConditions}. Note that taking the $r_+\rightarrow0$ limit of the above yields the boson star fields of {\S}\ref{PertBS} as it ought to.

\subsubsection{Near Region}\label{NearRegion}

Since we are constructing solutions perturbatively we are assuming low energies and angular momenta, so by construction $r_+\ll\ell$.  Just as in our asymptotic expansion we expanded in the dimensionless parameter $\epsilon\ll1$, near the horizon we expand in the dimensionless parameter $r_+/\ell\ll1$.  To accomplish this, we switch to the radial coordinate $z\equiv \ell r /r_+ $ such that the horizon is located at $z=\ell$.  This ensures that $z/\ell\ge1$ is always large with respect to $r_+/\ell\ll1$ so we can safely expand our fields in powers of $r_+/\ell$ without needing to worry about competing effects at the same order.  Note that this was also achieved in our asymptotic expansion since $r/\ell\gg1$ and the expansion parameter was $\epsilon\ll1$.  Thus, we perform our double expansion in the inner region using $z$ as our radial coordinate.

The gravitational and scalar field expansions are given by
\begin{equation}
F^{in}(z,\epsilon,r_+) = \sum_{i=0}^m \sum_{j=0}^n \tilde{F}^{in}_{2j, 2i}(z) \epsilon^{2j} \left(\frac{r_+}{\ell}\right)^{2i}, \quad\quad\quad
\Pi^{in}(z,\epsilon,r_+) = \sum_{i=0}^m \sum_{j=0}^n \tilde{\Pi}^{in}_{2j+1, 2i+1}(z) \epsilon^{2j+1} \left(\frac{r_+}{\ell}\right)^{2i+1} \label{eq:NearFieldExpansion}
\end{equation}
where $F^{in}=\{f^{in},g^{in},h^{in},\Omega^{in}\}$ are the metric functions in the near-region.  Recall, however, that the frequency $\omega$ is still given by Eq. (\ref{eq:BHomegaExpansion}) since it is defined globally.

Again, we start at order $\epsilon^0$ with the Myers-Perry-AdS solution, with gravitational fields $\tilde{F}^{in}_{0,2i}$.  We insert this into the equations of motion, appropriately transformed to equations of $z$, add a non-trivial scalar field by expanding to order $\epsilon$ and solve the equations order by order in $r_+/\ell$.  Once all orders in $r_+/\ell$ have been calculated up to the desired cutoff, the matching conditions must be imposed before continuing up the perturbative ladder.  A discussion of this procedure is postponed to {\S}\ref{MatchingConditions}.  Next, the scalar field at order $\epsilon$ sources the gravitational fields at order $\epsilon^2$.  These have to be solved order by order in $r_+/\ell$ and matched to the far region expansions before the process is continued.  The result of this procedure, with the matching conditions already imposed, up to ${\cal O}(\epsilon^a(r_+/\ell)^b)$ such that $a+b\le6$, is given by
\begin{align}
f^{in}(z)={}&\left[1-\frac{\ell^{n-1}}{z^{n-1}}+\left\{\frac{z^2}{\ell^2}-\big((n+2)^2+1\big)\frac{\ell^{n-1}}{z^{n-1}}+(n+2)^2\frac{\ell^{n+1}}{z^{n+1}}\right\}\frac{r_+^2}{\ell^2}\right.\label{eq:fNearRegion}\\
&\left.+\bigg(2(n+2)\omega_{n;0,2}\ell+(n+2)^2\big((n+2)^2-1\big)\bigg)\left(\frac{\ell^{n+1}-\ell^{n-1}z^2}{z^{n+1}}\right)\frac{r_+^4}{\ell^4}+{\cal O}\left(\frac{r_+^6}{\ell^6}\right)\right]\nonumber\\
&+\epsilon^2\left[\frac{r_+^2}{\ell^2}\frac{\ell^{n-1}}{z^{n-1}}\left(\int_{1}^{z/\ell}{\left(\log\left[1-\frac{1}{x^2}\right]f^{in}_{n;2,2}(x)dx\right)}-\log\left[1-\frac{\ell^2}{z^2}\right]\int_{1}^{z/\ell}{f^{in}_{n;2,2}(x)dx}\right)+{\cal O}\left(\frac{r_+^4}{\ell^4}\right)\right]\nonumber\\
&+\epsilon^4{\cal O}\left(\frac{r_+^2}{\ell^2}\right)\nonumber\\
g^{in}(z)={}&\bigg[1-(n+2)^2\frac{\ell^{n+1}}{z^{n+1}}\frac{r_+^2}{\ell^2}+\bigg\{-\bigg(2(n+2)\omega_{n;0,2}\ell+(n+2)^2\big((n+2)^2-1\big)\bigg)\frac{\ell^{n+1}}{z^{n+1}}\label{eq:gNearRegion}\\
&+(n+2)^4\frac{\ell^{2n+2}}{z^{2n+2}}\bigg\}\frac{r_+^4}{\ell^4}+{\cal O}\left(\frac{r_+^6}{\ell^6}\right)\bigg]+\epsilon^2\left[-\frac{2}{n}+\frac{r_+^2}{\ell^2}\left\{\frac{4\Gamma^4\big[\frac{n}{n-1}\big]}{n\Gamma^2\big[\frac{n+1}{n-1}\big]}\int_1^{z/\ell}{x\left(\big({P_{\frac1{n-1}}}[2x^{n-1}-1]\big)'\right)^2dx}\right.\right. \nonumber\\
&\left.+K^{g}_{n;2,2}+g^{in}_{n;2,2}\bigg\}+{\cal O}\left(\frac{r_+^4}{\ell^4}\right)\right]+\epsilon^4\left[{-\frac{g_{n;4,0}(0)}{\ell^{3n+7}}+\cal O}\left(\frac{r_+^2}{\ell^2}\right)\right]\nonumber\\
h^{in}(z)={}&\bigg[1+(n+2)^2\frac{\ell^{n+1}}{z^{n+1}}\frac{r_+^2}{\ell^2}+\bigg(2(n+2)\omega_{n;0,2}\ell+(n+2)^2\big((n+2)^2-1\big)\bigg)\frac{\ell^{n+1}}{z^{n+1}}\frac{r_+^4}{\ell^4}+{\cal O}\left(\frac{r_+^6}{\ell^6}\right)\bigg]\nonumber\\
&+\epsilon^2\left[h_{n;2,2}^{in}\frac{r_+^2}{\ell^2}+{\cal O}\left(\frac{r_+^4}{\ell^4}\right)\right]+\epsilon^4{\cal O}\left(\frac{r_+^2}{\ell^2}\right),\label{eq:hNearRegion}\\
\ell\Omega^{in}(z)={}&\bigg[(n+2)\frac{\ell^{n+1}}{z^{n+1}}+\left\{\ell\omega_{n;0,2}\frac{\ell^{n+1}}{z^{n+1}}+(n+2)^3\big(z^{n+1}-\ell^{n+1}\big)\frac{\ell^{n+1}}{z^{2n+2}}\right\}\frac{r_+^2}{\ell^2}+\bigg\{\ell\omega_{n;0,4}\frac{\ell^{n+1}}{z^{n+1}}\label{eq:OmegaNearRegion}\\
&+\big(z^{n+1}-\ell^{n+1}\big)(n+2)^2\bigg(\bigg(3\ell\omega_{n;0,2}+(n+2)\big((n+2)^2-1\big)\bigg)\frac{\ell^{n+1}}{z^{2n+2}}+(n+2)^3\frac{\ell^{2n+2}}{z^{3n+3}}\bigg)\bigg\}\frac{r_+^4}{\ell^4}\nonumber\\
&+{\cal O}\left(\frac{r_+^6}{\ell^6}\right)\bigg]+\epsilon^2\bigg[\ell\omega_{n;2,0}+\big(\ell\Omega_{n;2,0}(0)-\ell\omega_{n;2,0}\big)\left(1-\frac{\ell^{n+1}}{z^{n+1}}\right)+\frac{r_+^2}{\ell^2}\bigg\{\omega_{n;2,2}+\int_{1}^{z/\ell}\bigg[\frac{K^{\Omega}_{n;2,2}}{x^{n+2}}\nonumber\\
&+\frac{1}{x^{n+2}}\int_{1}^{x}\left[\frac{4(n+2)}{y^{n+2}(y^{n-1}-1)}\left(\Omega^{in}_{n;2,2}+y^{2n}\bigg(\frac{n+3}{2}-y^{n+1}\bigg)\frac{\Gamma^4\big[\frac{n}{n-1}\big]}{\Gamma^2\big[\frac{n+1}{n-1}\big]}\big(P_{\frac1{n-1}}[2y^{n-1}-1]\big)^2\right.\right.\nonumber\\
&\left.\left.-\frac{(n^2-1)y^{n+1}}{4(y^{n-1}-1)}\int_1^y{f^{in}_{n;2,2}(w)dw}\right)\right]dy\bigg]dx\bigg\}+{\cal O}\left(\frac{r_+^4}{\ell^4}\right)\bigg]+\epsilon^4\bigg[\ell\omega_{n;4,0}
\nonumber\\
&+\big(\ell\Omega_{n;4,0}(0)-\ell\omega_{n;4,0}\big)\left(1-\frac{\ell^{n+1}}{z^{n+1}}\right)+{\cal O}\left(\frac{r_+^2}{\ell^2}\right)\bigg],\nonumber\\
\Pi^{in}(z)={}&\epsilon\left[\frac{\Gamma^2\big[\frac{n}{n-1}\big]}{\Gamma\big[\frac{n+1}{n-1}\big]}P_{\frac{1}{n-1}}\bigg[2\frac{z^{n-1}}{\ell^{n-1}}-1\bigg]\frac{r_+}{\ell}+\frac{r_+^3}{\ell^3}\bigg\{K^{\Pi}_{n;1,3}P_{\frac{1}{n-1}}\left[\frac{2z^2}{\ell^2}-1\right]+\frac{2\Gamma^2\big[\frac{n}{n-1}\big]}{(n-1)\Gamma\big[\frac{n+1}{n-1}\big]}\times\right.\label{eq:PiNearRegion}\\
&\times\left[Q_{\frac{1}{n-1}}\bigg[\frac{2z^{n-1}}{\ell^{n-1}}-1\bigg]\int_1^{z/\ell}{P_{\frac1{n-1}}\big[2x^{n-1}-1\big]\frac{P_{\frac1{n-1}}\big[2x^{n-1}-1\big]\tilde{s}_n(x)+nP_{\frac{n}{n-1}}\big[2x^{n-1}-1\big]\hat{s}_n(x)}{x^3\left(\sum_{j=0}^{\frac{n-3}2}{x^{2j}}\right)^2}dx}\right.\nonumber\\
&\left.-P_{\frac1{n-1}}\bigg[\frac{2z^{n-1}}{\ell^{n-1}}-1\bigg]\int_1^{z/\ell}{Q_{\frac{1}{n-1}}\big[2x^{n-1}-1\big]\frac{P_{\frac1{n-1}}\big[2x^{n-1}-1\big]\tilde{s}_n(x)+nP_{\frac{n}{n-1}}\big[2x^{n-1}-1\big]\hat{s}_n(x)}{x^3\left( \sum_{j=0}^{\frac{n-3}2}{x^{2j}}\right)^2}dx}\right]\bigg\}\nonumber\\
&\left.+{\cal O}\left(\frac{r_+^5}{\ell^5}\right)\right]+\epsilon^3\bigg[K^{\Pi}_{n;3,1}P_{\frac{1}{n-1}}\left[2\frac{z^{n-1}}{\ell^{n-1}}-1\right]\frac{r_+}{\ell}+{\cal O}\left(\frac{r_+^3}{\ell^3}\right)\bigg]+\epsilon^5\left[{\cal O}\left(\frac{r_+}{\ell}\right)\right].\nonumber
\end{align}
where the constants $K^{g}_{n;2,2}$, $K^{\Omega}_{n;2,2}$ and $K^{\Pi}_{n;1,3}$ are calculated in {\S}\ref{MatchingConditions} and the constants $K^{\Pi}_{n;3,1}$ in Eq. (\ref{eq:PiNearRegion}) are given by 
\begin{equation}
K^{\Pi}_{3;3,1}=\frac{55\pi}{288},\quad\quad K^{\Pi}_{5;3,1}=\frac{1067\Gamma^2\big[\frac{5}{4}\big]}{900\sqrt{\pi}},\quad\quad K^{\Pi}_{7;3,1}=\frac{6403\Gamma^2\big[\frac{7}{6}\big]}{14112\Gamma\big[\frac{4}{3}\big]},\quad\quad K^{\Pi}_{9;3,1}=\frac{1104601\Gamma^2\big[\frac{9}{8}\big]}{3175200\Gamma\big[\frac{5}{4}\big]}
\end{equation}
for $n=3,\>5,\>7,\>9$ respectively.  The functions $P_\nu[y]$ and $Q_\nu[y]$ are the Legendre functions of the first and second kind, respectively, the functions $\tilde{s}_n(x)$ are given in Appendix B and
\begin{equation}
\hat{s}_n(x)=x^{n+1}\left(\displaystyle \sum_{j=0}^{\frac{n-3}{2}}{(j+1)x^{n-3-2j}}\right)+(n+2)^2\left(\displaystyle \sum_{j=0}^{\frac{n-3}{2}}{(j+1)x^{2j}}\right).
\end{equation}

There are a few discrepancies with Ref. \cite{Dias:2011at} (the $n=3$ case) that warrant mentioning.  In the order $\epsilon(r_+/\ell)^3$ term of Eq. (\ref{eq:PiNearRegion}), we used the matching condition outlined in the next subsection and determined that the constants $K^{\Pi}_{n;1,3}\ne0$, contrary to what is assumed  in \cite{Dias:2011at}.  Next, although $Q_{\frac{1}{n-1}}[y]$ is complex for $y>1$, it can be explicitly verified both in the near horizon region and in the large $z$ limit, the explicit procedure for which is discussed in the next subsection, that the imaginary part in this term cancels for our solution whereas it does not cancel for the solution quoted in \cite{Dias:2011at}.  Furthermore, the authors found no order $\epsilon^3(r_+/\ell)$ term in Eq. (\ref{eq:PiNearRegion}) whereas we find that $K^{\Pi}_{n;3,1}\ne0$.  Despite these discrepancies  in the near-horizon scalar field, the stress energy tensor is unaffected at the perturbative   order we are probing.  Similarly, the First Law is also satisfied to the appropriate perturbative order for
both our solutions and those of Ref. \cite{Dias:2011at}, so these discrepancies can only be distinguished at higher orders; we postpone a discussion of the thermodynamics of our
solutions until {\S}\ref{Thermo}.
Finally, although our solutions for $f^{in}$ and $g^{in}$ at order $\epsilon^2(r_+/\ell)^2$ look different than the result in \cite{Dias:2011at}, it can be verified that our solution for $n=3$ agrees term by term in the near horizon expansion.

\subsubsection{Matching Region}\label{MatchingConditions}

A crucial and often difficult step in this analysis involves matching the near region solution to the far region solution in order to ensure a valid solution everywhere.  The heuristic procedure is as follows: take a small-$r$ expansion of the outer region fields, i.e. expand $F^{out}$ around $r=0$, and match these at each order in $\epsilon$ and $r_+$ to the large-$z$ expansion of the inner fields, i.e. an expansion of $F^{in}$ around $z=\infty$.  Such a matching takes place in an area between the two regions, where $r_+\ll r\ll\infty$.  This then fixes the remaining arbitrary constants of integration in the fields as well as fixes the expansion parameters for $\omega$.  When taking the large $z$ limit of the inner region fields, one must first pick an order in $\epsilon$, take the large $z$ limit at that order, transform the limit back to the original radial coordinate, $r=r_+z/\ell$, and finally expand the result in powers of $r_+/\ell$.  In principle, this is a straightforward procedure; in practice it can be subtle and difficult.  This is best illustrated with an explicit example.  

Consider the large $z$ expansion of the order $\epsilon^2$ term of $g^{in}(z)$ in $n=3$.  The difficulty arises because we must take the large $z$ limit of a function that is defined via an integral of the form
\begin{equation}
\int_{1}^{z/\ell}{f(x)dx}\label{eq:ConvergentIntegral}
\end{equation}
for some function $f(x)$.  Now, if (\ref{eq:ConvergentIntegral}) converges as we send $z\rightarrow\infty$ then this integral is easy enough to perturbatively compute.  Indeed
\begin{equation}
\int_{1}^{z/\ell}{f(x)dx}=\int_{1}^{\infty}{f(x)dx}-\int_{z/\ell}^{\infty}{f(x)dx}
\end{equation}
where now $\int_{1}^{\infty}{f(x)dx}$ is a constant that we can easily compute, albeit numerically, and the second term can be evaluated by taking a Taylor series expansion of $f(x)$ near $\infty$.  A problem arises, however, if the integral (\ref{eq:ConvergentIntegral}) does \emph{not} converge.  This is true of the integral appearing in our example and is, in fact, generically true for the \emph{majority} of the inner region fields above defined through integrals.  We therefore must find a way to correctly take the large $z$ limit of these integrals.

Motivated by the above discussion for convergent integrals, if $\int_{1}^{\infty}{f(x)dx}$ diverges, we define $Div(x)$ as the sum of the terms which cause the integral to diverge.  That is, we Taylor expand $f(x)$ at infinity and collect the terms whose integral diverges and call that $Div(x)$.  Then, by definition, $\int_{1}^{\infty}{(f(x)-Div(x))dx}$ converges and the above consideration is valid.  This allows us to rewrite our original integral as
\begin{equation}
\int_{1}^{z/\ell}{f(x)dx}=\int_{1}^{\infty}{\big(f(x)-Div(x)\big)dx}-\int_{z/\ell}^{\infty}{\big(f(x)-Div(x)\big)dx}+\int_{1}^{z/\ell}{Div(x)dx} \label{eq:largezIntegral}
\end{equation}
Now $\int_{1}^{\infty}{\big(f(x)-Div(x)\big)dx}$ is a numerical constant, the second integral can be evaluated by Taylor expanding $f(x)-Div(x)$ near infinity and since $Div(x)$ is a simple power series in $x$, the last integral has an exact analytic expression.  Note that we are free to include extra convergent terms in the definition of $Div(x)$ since such terms will consequently disappear from to the first two integrals of (\ref{eq:largezIntegral}), not changing the end result.

Applying this technique to the order $\epsilon^2$ term of $g^{in}(z)$, we identify $f(x)=\frac{\pi^2}{12}x\left((P_{\frac12}[2x^2-1])'\right)^2$.  Expanding $f(x)$ around infinity, we find 
$$
Div(x)=\frac{4x}{3}+\frac{2}{3x}-\frac{\log[x]}{2x^3} \qquad 
\int_{1}^{\infty}{\big(f(x)-Div(x)\big)dx}=-10.3749804991685... \equiv C_1
$$
 up to 15 digits of precision.  We now have the large $z$ limit of $f(x)$ since the other integrals are straightforward to compute.  Let us now detail how the matching condition is employed.

We must take the large $z$ limit of the entire order at $\epsilon^2$.  That is, we must expand
\begin{equation}
-\frac{2}{3}+\frac{r_+^2}{\ell^2}\left\{\int^{z/\ell}_{1}{f(x)dx}+K^{g}_{3;2,2}+g^{in}_{3;2,2}\right\}\label{eq:gIn32Match}
\end{equation}
at large $z$.  Next, we transform back to the coordinate $r=zr_+/\ell$ and series expand in powers of $r_+/\ell$.  The result for (\ref{eq:gIn32Match})  up to order $(r_+/\ell)^2$ is
\begin{equation}
g^{in}_{\epsilon^2}\rightarrow-\frac{2}{3}+\frac{2r^2}{3\ell^2}+\frac{r_+^2}{\ell^2}\left\{-\frac{19}{24}+C_1+K^{g}_{3;2,2}-\frac{2}{3}\log\left[\frac{r_+}{\ell}\right]+\frac{2}{3}\log\left[\frac{r}{\ell}\right]+{\cal O}(r^2)\right\}+{\cal O}\left(\frac{r_+^4}{\ell^4}\right).\label{eq:gInMatch}
\end{equation}
This is now to be matched with the small $r$ limit of the order $\epsilon^2$ term of $g^{out}(r)$.  Direct calculation using the fields in Appendix B (with the constant of integration, $C_2$, not yet fixed) yields
\begin{equation}
g^{out}_{\epsilon^2}\rightarrow-\frac{2}{3}+\frac{2r^2}{3\ell^2}+\frac{r_+^2}{\ell^2}\left\{-\frac{C_2\ell^4}{4r^4}+\frac{C_2\ell^2}{r^2}+\frac{C_2}{2}+3C_2\log\left[\frac{r}{\ell}\right]-\frac{83}{9}+\frac{2}{3}\log\left[\frac{r}{\ell}\right]+{\cal O}(r^2)\right\}+{\cal O}\left(\frac{r_+^4}{\ell^4}\right).\label{eq:gOutMatch}
\end{equation}
The order $(r_+/\ell)^0$ terms cancel so we continue to the order $(r_+/\ell)^2$ terms: Eq. (\ref{eq:gOutMatch}) has a term of order $r^{-4}$ while Eq. (\ref{eq:gInMatch}) does not, meaning that we must set $C_2=0$.  The $\log[r/\ell]$ terms cancel and we are left with a condition on the constant $K^{g}_{3;2,2}$.  Explicit calculation reveals
\begin{equation}
K^{g}_{3;2,2}=-\frac{607}{72}-C_1+\frac{2}{3}\log\left[\frac{r_+}{\ell}\right]
\end{equation}
which has a numerical value of $1.94442...+\frac{2}{3}\log\left[\frac{r_+}{\ell}\right]$.  This result differs from the value of $\frac{635}{504}+\frac{7}{8}\log[2]+\frac{2}{3}\log\left[\frac{r_+}{\ell}\right]\approx1.86642...+\frac{2}{3}\log\left[\frac{r_+}{\ell}\right]$ in Ref. \cite{Dias:2011at} well beyond the margins of numerical error.  This discrepancy arises because in Ref.~\cite{Dias:2011at} an approximation scheme was used to evaluate the large-$z$ limit of these integrals~\cite{Dias:private}, whereas our results are exact.

Following the procedure outlined above, in $n=5,7,9$ we find the corresponding constants to be
\begin{equation}
K^g_{5;2,2}=9.020251036253986, \quad\quad K^g_{7;2,2}=8.747040877884002,\quad\quad K^g_{9;2,2}=9.031578102169204
\end{equation}
Similar considerations show that $K^{\Pi}_{n;1,3}$ are given by
\begin{align}
K^{\Pi}_{3;1,3}={}&-14.36212300918522-\frac{21\pi}{8}\log\left[\frac{r_+}{\ell}\right]\quad\quad K^{\Pi}_{5;1,3}=-5.48986673419516\\
K^{\Pi}_{7;1,3}={}&-7.87633500695327\quad\quad\quad\quad\quad\quad\quad\quad\>\>\>\quad K^{\Pi}_{9;1,3}=-9.11278197442749
\end{align}

Evaluation of the constants $K^{\Omega}_{n;2,2}$ requires further subtle considerations since the corresponding field is defined via a series of nested integrals.  The procedure outlined above fails under such conditions.  This is remedied by treating the inner integrals first by Taylor expanding the integrand at infinity and near the horizon, while keeping a sufficient number of terms in each series such that the two match-up at some intermediate radius.  This integral can then be evaluated directly, with the answer then becoming the integrand of the next integral.  This integrand can then be expanded in the same way and its answer used as the integrand of the next integral in the nest.  Finally, the last integral in the nest can be treated as discussed above to obtain the large $z$ limit of the field.  This rather tedious procedure yields
\begin{align}
K^{\Omega}_{3;2,2}&=64.6979...-10\log\left[\frac{r_+}{\ell}\right] \quad\quad\quad K^{\Omega}_{5;2,2}=306.637...\\
&K^{\Omega}_{7;2,2}=958.528...\quad\quad\quad K^{\Omega}_{9;2,2}=2451.23...\nonumber
\end{align}
where we have kept only 6 digits of precision due to the numerical error introduced by the repeated matching of Taylor series.  We again have a discrepancy well beyond the bounds of numerical error with the results of \cite{Dias:2011at}, which found $K^{\Omega}_{3;2,2}=87.5209...-10\log\left[\frac{r_+}{\ell}\right]$, again because of the approximation procedure used \cite{Dias:private}.

\section{Thermodynamics and Physical Properties}\label{Thermo}

Since our solution is invariant under the Killing vector field (\ref{eq:KV}), it must satisfy the First Law of Thermodynamics, which follows from a Hamiltonian derivation of the first law \cite{Wald:1993ki}.  In this section we work out the thermodynamic quantities of our hairy black hole solution and verify that the first law holds in each dimension up to the appropriate perturbative order.  

\subsection{Asymptotic Charges}

In an asymptotically anti-de Sitter spacetime $(M,g_{ab})$ there is an ambiguity in defining ``the" asymptotic metric because certain metric functions diverge and thus there does not exist a smooth limit to the boundary.  To get around this subtlety, Penrose proposed a conformal completion $(\hat{M},\hat{g}_{ab})$, where $\hat{g}_{ab}=\tilde\Omega^2g_{ab}$ for some conformal factor $\tilde\Omega$, such that the boundary of $\hat{M}$ is reached by the smooth limit $\tilde\Omega\rightarrow0$.  In particular, by virtue of $\tilde\Omega$ vanishing (smoothly) on the boundary, $\nabla\tilde\Omega$ can be used as a radial direction near infinity and the subtleties of taking infinite distance limits in the physical metric $g_{ab}$ reduces to local differential geometry of fields in the conformal completion $\hat{g}_{ab}$ over finite distances.  Using the conformal completion with reflective boundary conditions,  Ashtekar and Das elucidated how to properly define conserved charges in asymptotically anti-de Sitter spacetimes \cite{Ashtekar:1999jx}, a procedure
that was extended to the rotating case in \cite{Das:2000cu}.
Furthermore, it has been pointed out that while there exist results in the literature in disagreement with the Ashtekar-Das formalism, such definitions of mass and angular momenta fail to satisfy the first law \cite{Gibbons:2004ai}.  We therefore restrict our attention to the Ashtekar-Das formalism for computing the mass and angular momentum, briefly detailing the procedure.

We start by introducing the conformal metric $\hat{g}_{ab}=\tilde\Omega^2g_{ab}$ where the physical metric is given in (\ref{eq:metric}) and the appropriate conformal factor is $\tilde\Omega=1/r$.  The conformal metric explicitly looks like
\begin{equation}
d\hat{s}^2=-\tilde{f}gdt^2+\frac{d\tilde\Omega^2}{\tilde{f}}+h\big(d\chi+A_idx^i-\Omega dt\big)^2+g_{ij}dx^idx^j\label{eq:ConformalMetric}
\end{equation}
where $g,h$ and $\Omega$ are our previous (physical) metric functions written in terms of the conformal radial coordinate, $\tilde\Omega$, and $\tilde{f}=\tilde\Omega^2f$ now admits a smooth, finite limit to the conformal boundary, $\cal I$, defined by $\tilde\Omega=0$.  The Weyl tensor, $\hat{C}_{abcd}$, of the conformal metric (\ref{eq:ConformalMetric}) vanishes on $\cal I$ but ${\bf K}_{abcd}\equiv\displaystyle \lim_{\rightarrow{\cal I}}\tilde\Omega^{3-D}\hat{C}_{abcd}$ admits a smooth limit, the electric part of which, ${\cal E}_{ab}$, is used to define conserved quantities.  Using the transformation properties of the Weyl  and Ricci tensors under conformal rescalings and imposing the Einstein equations, it can be shown that ${\cal E}_{ab}$ satisfies the continuity equation \cite{Ashtekar:1999jx,Das:2000cu}
\begin{equation}
D^d{\cal E}_{md}=-8\pi(D-3){\bf T}_{ab}n^a{h^b}_m\label{eq:ContinuityEquation}
\end{equation}
where $n_a=\partial_a\tilde\Omega$ is the normal vector to $\cal I$, $D_a$ is the derivative operator compatible with the induced metric $h_{ab}$ on $\cal I$ and ${\bf T}_{ab}=\displaystyle\lim_{\rightarrow{\cal I}}{\tilde\Omega^{2-D}T_{ab}}$ admits a smooth limit where $T_{ab}$ is the matter stress tensor in Eq.  (\ref{eq:Tab}).  If there is no net flux of the matter stress tensor on $\cal I$, then the corresponding charge is conserved.  

To set this up explicitly, the electric part of the Weyl tensor takes the form ${\cal E}_{ab}\hat{=}\ell^2{\bf K}_{ambn}n^mn^n$ where the symbol $\hat=$ is used to denote equality on $\cal I$.  We also have that $n_a$ is spacelike, which means $\cal I$ is timelike.  Consider now a $t$=constant slice of $\cal I$: this is a $D-2$ dimensional spacelike hypersurface, $\Sigma$, with induced metric $\gamma$ and normal vector $t_a=\partial_a t$.  For any conformal Killing vector, $\xi^a$, the continuity equation (\ref{eq:ContinuityEquation}) leads to the conserved charge
\begin{equation}
Q_{\xi}=\pm\frac{1}{8\pi(n-1)}\int_{\Sigma}{{\cal E}_{ab}\xi^{a}t^{b}\mathrm{d}\Sigma}
\end{equation}
where the $\pm$ sign corresponds to a timelike/spacelike conformal Killing vector respectively and $n=D-2$ is not to be confused with $n_a$.

Although our solution is invariant under a single Killing field, both $\partial_t$ and $\partial_\chi$ are asymptotic Killing vectors since the scalar field vanishes asymptotically.  Note that the vanishing of the scalar field at infinity also ensures there will be no net flux of matter fields on the conformal boundary, and so the charges defined above are strictly conserved.   $\partial_t$ and $\partial_\chi$ are then the desired conformal Killing vectors and they lead to a conserved energy and angular momentum respectively.  A direct calculation of the conserved charges yields
\begin{equation}
E=\frac{(n+1)\pi^{\frac{n-1}{2}}\ell^{n-1}}{16\left(\frac{n+1}{2}\right)!}\left((n+1)C_h-nC_f\right)
\end{equation}
\begin{equation}
J=\frac{(n+1)^2\pi^{\frac{n-1}{2}}\ell^{n}C_\Omega}{16\left(\frac{n+1}{2}\right)!}
\end{equation}
where $C_f, C_h$ and $C_\Omega$ are the leading order corrections to the asymptotic fields $f, h$ and $\Omega$ appearing in the boundary conditions (\ref{eq:AsymBC}).  Using the far region field expansions (\ref{eq:fFarRegion}), (\ref{eq:hFarRegion}) and (\ref{eq:OmegaFarRegion}), along with the catalogued fields in Appendix B, these coefficients are easily obtained.  The resulting perturbative expansions of the energy and angular momentum charges for $n=3,5,7,9$ are catalogued in Appendix A.

\subsection{Near Horizon Quantities}

Boson stars are horizonless geometries so their thermodynamics are completely governed by the above subsection.  For the case of black holes, the presence of the horizon introduces a temperature and an entropy which also enter the first law.  We now wish to find these quantities.

The norm of the Killing vector (\ref{eq:KV}) is $\left|K\right|^2=-fg+r^2(\omega-\Omega)^2$, which is null at the horizon by virtue of the condition $\Omega_H=\omega$ and the fact that $f$ vanishes at the horizon. The Killing vector under which our solution is invariant is therefore tangent to the generators of the horizon.  The event horizon is therefore also a Killing horizon and thus has a temperature $T_H=\frac{\kappa}{2\pi}$.  For a metric of the form (\ref{eq:metric}), a straightforward calculation yields
\begin{equation}
T_H=\frac{1}{4\pi}{f'(r_+)\sqrt{g(r_+)}} \label{eq:Temperature}
\end{equation}
Furthermore, any solution to Einstein's equations with or without cosmological constant obeys the Bekenstein-Hawking area-entropy law.  Spatial sections of the horizon have an induced metric of a squashed $n$-sphere
\begin{equation}
ds_H^2=r^2(h\big(d\chi+A_idx^i)^2+g_{ij}dx^idx^j)\bigg|_{r_+}
\end{equation}
so the entropy takes the form
\begin{equation}
S=\frac{A_n}{4}r_+^n\sqrt{h(r_+)}\label{eq:Entropy}
\end{equation}
where $A_n$ is the area of a (round) unit $n$-sphere.  Now, the near region field expansions (\ref{eq:fNearRegion}), (\ref{eq:hNearRegion}) and (\ref{eq:OmegaNearRegion}), along with the fields catalogued in Appendix B yield the perturbative expansions of (\ref{eq:Temperature}) and (\ref{eq:Entropy}).  These expansions for the entropy are also catalogued in Appendix A for $n=3,5,7,9$.

With the thermodynamic charges and potentials catalogued in Appendix A, it is straightforward to verify that the first law holds up to the appropriate perturbative order in each dimension.  For boson stars, obtained in the $r_+\rightarrow0$ limit, the first law takes the form
\begin{equation}
\mathrm{d}E=\omega \mathrm{d}J
\end{equation}
while for the hairy black holes, the first law takes the form
\begin{equation}
\mathrm{d}E=T_H\mathrm{d}S+\omega \mathrm{d}J.
\end{equation}

Upon examining the thermodynamic quantities in Appendix A, we see that the angular momentum is primarily carried by the scalar field.  Furthermore, the scalar field carries more and more of the angular momentum as the space-time dimension increases: the leading terms in the perturbative expansions are due to the scalar field at orders $\epsilon^2,...,\epsilon^{n-1}$ and the next terms are a mixture of scalar field and black hole at orders $r_+^p\epsilon^q$ where $p+q=n+1$.  Similarly, for $n=3$ the energy has contributions at the same perturbative order from the black hole and the scalar field since the leading terms in $E_3$ are of order $\epsilon^2$ and $r_+^2$.  However, for $n>3$ the majority of the energy is carried again by the scalar field: the leading terms in $E_n$ are at orders $\epsilon^2,...,\epsilon^{n-3}$ and the next terms are at orders $r_+^{n-1}$ and $\epsilon^{n-1}$.  This is a rather striking feature of the perturbative regime of these hairy black holes: in $n=3$ the thermodynamics are governed both by the scalar field and the black hole while in $n>3$ the thermodynamics are dominated by the scalar field.

\section{Discussion}\label{Discussion}

We have constructed perturbative solutions describing asymptotically AdS rotating hairy black holes and boson stars in dimensions $D=5,7,9,11$.  Apart from several technical discrepancies, we are in general agreement with previous
results \cite{Dias:2011at} for the $D=5$ case.  Such solutions describe lumpy massless scalar hair co-rotating with a black hole and are invariant under a single Killing field.  This is made possible by a particular choice of scalar field ansatz, whose stress tensor shares the same symmetries as the metric.  These are the first known examples of asymptotically AdS black holes that are invariant under one Killing vector; all previous AdS black holes were stationary and hence had at least two Killing vector fields.

The hairy black hole solutions constructed herein describe a two-parameter family of solutions, characterised by $r_+$ and $\epsilon$, which is connected to the two-parameter family of Myers-Perry-AdS black holes, characterised by $r_+$ and $a$; the introduction of the scalar field and requiring invariance under a single Killing vector fixes the hairy black hole angular momentum, leading to a two-parameter family instead of three.  The Myers-Perry-AdS family is obtained in the limit of vanishing scalar hair.  As pointed out in \cite{Dias:2011at}, sufficiently low frequency perturbations in the Myers-Perry-AdS solution will lead to a super-radiant instability where the amplitude of the perturbation will grow at the expense of the black hole rotation energy.  During the instability, there are multiple frequency parameters corresponding to the various perturbation modes and the solution therefore is not invariant under any Killing fields.  The end result of this instability, at least in the perturbative regime, will only have the lowest frequency mode present since higher frequency modes will get eaten by the horizon as the lower frequency modes continue to mine rotational energy via super-radiance.   The hairy black hole solution considered herein, then, is the end result of a super-radiant instability of the 
odd-dimensional Myers-Perry-AdS black hole; the condition that the angular velocity of the scalar field matches the angular velocity of the horizon is a statement that the mode has extracted all the rotational energy it can and super-radiant scattering is no longer possible for that mode.  Furthermore, these considerations motivate our only considering the ground state radial mode (\ref{eq:Piepsilon}) of the scalar hair and ignoring all higher radial modes since this corresponds to the lowest frequency mode.

Our two-parameter family of solutions further admits a one-parameter family limit that describes rotating boson stars, which are horizonless geometries with a rotating clump of scalar field condensate.  This (continuous) limit corresponds to $r_+\rightarrow0$, which means that a perturbative hairy black hole is interpreted as a small black hole being added to the center of a boson star.  This interpretation is explicitly clear by the double field expansion in powers of $\epsilon$ and $r_+$.  One can first perform an expansion in powers of $\epsilon$ to obtain the boson star solution, then expand in powers of $r_+$ to obtain the hairy black hole: the second expansion corresponds to adding a black hole to the center of the boson star.  It is possible to add a black hole to the center of a rotating boson star only if $\omega=\Omega_H$, which is explicitly demanded by the equations of motion, so that there is no net flux of scalar field across the horizon.

The present work is a step in the direction of further understanding AdS solutions with one Killing vector, but it is far from complete.  We have only considered the perturbative regime where energy and angular momenta are small by construction.  In \cite{Dias:2011at}, non-perturbative results were also investigated in $D=5$ and a rather rich and interesting thermodynamic structure was unveiled.  The drawback is that one must construct solutions numerically because of the highly non-linear and coupled equations of motion.  The benefit is that the physics is much more interesting; we refer the reader to Ref. \cite{Dias:2011at} for a full discussion.  As noted in {\S}\ref{Setup}, there are terms in the equations of motion which are absent in $D=5$ and it is unclear from our perturbative analysis whether these terms have any interesting physical significance to higher dimensional hairy black holes.  The only way to uncover this potential new physics would be to numerically construct hairy black holes in $D=7,9,11$ and compare them to the numerical solutions in $D=5$.  This is not a light undertaking and it is currently left for future considerations.

It is also unclear from the present construction whether it is possible to construct asymptotically AdS hairy black holes with one Killing field in three-dimensions; such a solution would correspond to a BTZ black hole with lumpy scalar hair.  This would be a desirable solution to have since $D=3$ is the quintessential playground for investigations of quantum gravity.  However, since there does not exist a non-trivial Hopf-fibration of the one-sphere the prescription used in this paper is inappropriate to search for solutions in three-dimensions.  This is related to the $D\ge 5$ Myers-Perry black holes being disjoint from the BTZ black holes.  One must then use a different technique to construct three-dimensional hairy black holes, which are presumably analogously connected to the BTZ black hole in the limit of vanishing scalar hair.  Likewise, there is nothing physically preventing analogous even-dimensional hairy black hole solutions from existing so it should be possible to repeat this analysis in even dimensions.  The hurdle in this respect will be finding the appropriate form of the scalar fields such that the stress tensor shares the symmetries of the metric.

Finally, it would be desirable to study these hairy black hole solutions from an AdS/CFT perspective to see what they correspond to in the dual gauge theory.  The interpretation of our solutions from a boundary gauge theory perspective is currently an open issue and certainly warrants future investigation.

\vskip .5cm
\centerline{ \bf Acknowledgements}
\vskip .2cm

This work was made possible by funds supplied by the Natural Sciences and Engineering Research Council of Canada.

\appendix

\section{Conserved Charges and Thermodynamic Quantities}

 In this Appendix  we catalogue the thermodynamic charges and potentials entering the first law
 in spacetime dimension $D=n+2$, for $n=3,5,7,9$.  The boson star quantities are obtained by the limit $r_+\rightarrow0$, except for the temperature, which exhibits the usual Schwarzschild-like divergence as $r_+\rightarrow0$; the boson star temperature is zero.  The constants $K^g_{n;2,2}$ appearing in the expressions for the temperatures are calculated in {\S}\ref{MatchingConditions}.

\subsection*{n=3}

\begin{align}
E_3={}&\frac{\pi\ell^2}{4}\left(\left[\frac{3}{2}\frac{r_+^2}{\ell^2}+39\frac{r_+^4}{\ell^4}+{\cal O}\left(\frac{r_+^6}{\ell^6}\right)\right]+\epsilon^2\left[\frac56+\frac{191}{48}\frac{r_+^2}{\ell^2}+{\cal O}\left(\frac{r_+^4}{\ell^4}\right)\right]+\epsilon^4\left[\frac{77951}{127008}+{\cal O}\left(\frac{r_+^2}{\ell^2}\right)\right]\right.\\
&+{\cal O}(\epsilon^6)\bigg)\nonumber\\
J_3={}&\frac{\pi\ell^3}{2}\left(\left[5\frac{r_+^4}{\ell^4}+{\cal O}\left(\frac{r_+^6}{\ell^6}\right)\right]+\epsilon^2\left[\frac1{12}+\frac{43}{96}\frac{r_+^2}{\ell^2}+{\cal O}\left(\frac{r_+^4}{\ell^4}\right)\right]+\epsilon^4\left[\frac{83621}{1270080}+{\cal O}\left(\frac{r_+^2}{\ell^2}\right)\right]\right.\\
&+{\cal O}(\epsilon^6)\bigg)\nonumber
\end{align}

\begin{align}
\ell\omega_3={}&\left[5-3\frac{r_+^2}{\ell^2}-\left\{\frac{959}{16}-\frac{3}{2}\log\left[\frac{r_+}{4\ell}\right]\right\}\frac{r_+^4}{\ell^4}+{\cal O}\left(\frac{r_+^6}{\ell^6}\right)\right]-\epsilon^2\left[\frac{15}{28}+\frac{4469}{840}\frac{r_+^2}{\ell^2}+{\cal O}\left(\frac{r_+^4}{\ell^4}\right)\right]\\
&-\epsilon^4\left[\frac{22456447}{35562240}+{\cal O}\left(\frac{r_+^2}{\ell^2}\right)\right]+{\cal O}(\epsilon^6)\nonumber\\
T_{H;3}={}&\frac{1}{\pi r_+}\left(\left[\frac12-\frac{71}{4}\frac{r_+^2}{\ell^2}-\frac{2665}{16}\frac{r_+^4}{\ell^4}+{\cal O}\left(\frac{r_+^6}{\ell^6}\right)\right]-\epsilon^2\left[\frac{1}{6}+\left\{\frac{59}{42}-\frac{K^g_{3;2,2}}4+\frac{\pi^2}{32}\right\}\frac{r_+^2}{\ell^2}+{\cal O}\left(\frac{r_+^4}{\ell^4}\right)\right]\right.\\
&\left.-\epsilon^4\left[\frac{101341}{508032}+{\cal O}\left(\frac{r_+^2}{\ell^2}\right)\right]\right)\nonumber\\
S_3={}&r_+^3\frac{\pi^2}{2}\left(\left[1+\frac{25}{2}\frac{r_+^2}{\ell^2}+\frac{1655}{8}\frac{r_+^4}{\ell^4}+{\cal O}\left(\frac{r_+^6}{\ell^6}\right)\right]+\epsilon^2\left[\frac{265}{84}\frac{r_+^2}{\ell^2}+{\cal O}\left(\frac{r_+^4}{\ell^4}\right)\right]+\epsilon^4{\cal O}\left(\frac{r_+^2}{\ell^2}\right)+{\cal O}(\epsilon^6)\right)
\end{align}

\subsection*{n=5}

\begin{align}
E_5={}&\frac{\pi^2\ell^4}{8}\left(\left[\frac{5}{2}\frac{r_+^4}{\ell^4}+125\frac{r_+^6}{\ell^6}+{\cal O}\left(\frac{r_+^8}{\ell^8}\right)\right]+\epsilon^2\left[\frac7{20}+\frac{3463}{480}\frac{r_+^4}{\ell^4}+{\cal O}\left(\frac{r_+^6}{\ell^6}\right)\right]+\epsilon^4\left[\frac{314018183}{2208492000}+{\cal O}\left(\frac{r_+^4}{\ell^4}\right)\right]\right.\nonumber\\
&+{\cal O}(\epsilon^6)\bigg)\\
J_5={}&\frac{\pi^2\ell^5}{8}\left(\left[21\frac{r_+^6}{\ell^6}+{\cal O}\left(\frac{r_+^8}{\ell^8}\right)\right]+\epsilon^2\left[\frac1{20}+\frac{529}{480}\frac{r_+^4}{\ell^4}+{\cal O}\left(\frac{r_+^6}{\ell^6}\right)\right]+\epsilon^4\left[\frac{327248543}{15459444000}+{\cal O}\left(\frac{r_+^4}{\ell^4}\right)\right]\right.\\
&+{\cal O}(\epsilon^6)\bigg)\nonumber
\end{align}

\begin{align}
\ell\omega_5={}&\left[7-10\frac{r_+^4}{\ell^4}-\left\{360+\frac{320\Gamma^2\big[\frac54\big]}{\Gamma^2\big[\frac{-1}4\big]}\right\}\frac{r_+^6}{\ell^6}+{\cal O}\left(\frac{r_+^8}{\ell^8}\right)\right]-\epsilon^2\left[\frac{514}{2145}+\frac{1438545023}{96621525}\frac{r_+^4}{\ell^4}+{\cal O}\left(\frac{r_+^6}{\ell^6}\right)\right]\\
&-\epsilon^4\left[\frac{60223029794867}{315024820110000}+{\cal O}\left(\frac{r_+^4}{\ell^4}\right)\right]+{\cal O}(\epsilon^6)\nonumber\\
T_{H;5}={}&\frac{1}{\pi r_+}\left(\left[1-\frac{95}{2}\frac{r_+^2}{\ell^2}-\frac{5985}{8}\frac{r_+^4}{\ell^4}+{\cal O}\left(\frac{r_+^6}{\ell^6}\right)\right]-\epsilon^2\left[\frac{1}{5}+\left\{\frac{10559}{2145}-\frac{K^g_{5;2,2}}2+\frac{2\Gamma^4\big[\frac{5}{4}\big]}{\pi}\right\}\frac{r_+^2}{\ell^2}+{\cal O}\left(\frac{r_+^4}{\ell^4}\right)\right]\right.\nonumber\\
&\left.-\epsilon^4\left[\frac{997904461}{5521230000}+{\cal O}\left(\frac{r_+^2}{\ell^2}\right)\right]\right)\\
S_5={}&r_+^5\frac{\pi^3}{4}\left(\left[1+\frac{49}{2}\frac{r_+^2}{\ell^2}+\frac{6447}{8}\frac{r_+^4}{\ell^4}+{\cal O}\left(\frac{r_+^6}{\ell^6}\right)\right]+\epsilon^2\left[\frac{19831}{4290}\frac{r_+^2}{\ell^2}+{\cal O}\left(\frac{r_+^4}{\ell^4}\right)\right]+\epsilon^4{\cal O}\left(\frac{r_+^2}{\ell^2}\right)+{\cal O}(\epsilon^6)\right)
\end{align}

\subsection*{n=7}

\begin{align}
E_7={}&\frac{\pi^3\ell^6}{24}\left(\left[\frac{7}{2}\frac{r_+^6}{\ell^6}+287\frac{r_+^8}{\ell^8}+{\cal O}\left(\frac{r_+^{10}}{\ell^{10}}\right)\right]+\epsilon^2\left[\frac9{70}+\frac{60551}{5600}\frac{r_+^6}{\ell^6}+{\cal O}\left(\frac{r_+^8}{\ell^8}\right)\right]+\epsilon^4\left[\frac{3302488311}{117937339520}\right.\right.\\
&\left.+{\cal O}\left(\frac{r_+^6}{\ell^6}\right)\right]+{\cal O}(\epsilon^6)\bigg)\nonumber\\
J_7={}&\frac{\pi^3\ell^7}{2}\left(\left[3\frac{r_+^8}{\ell^8}+{\cal O}\left(\frac{r_+^{10}}{\ell^{10}}\right)\right]+\epsilon^2\left[\frac1{840}+\frac{7039}{67200}\frac{r_+^6}{\ell^6}+{\cal O}\left(\frac{r_+^8}{\ell^8}\right)\right]+\epsilon^4\left[\frac{10188926113}{38211698004480}+{\cal O}\left(\frac{r_+^6}{\ell^6}\right)\right]\right.\nonumber\\
&+{\cal O}(\epsilon^6)\bigg)
\end{align}

\begin{align}
\ell\omega_7={}&\left[9-35\frac{r_+^6}{\ell^6}-\left\{2240+\frac{560\Gamma\big[\frac{-4}{3}\big]\Gamma^2\big[\frac76\big]}{\Gamma\big[\frac{4}{3}\big]\Gamma^2\big[\frac{-1}6\big]}\right\}\frac{r_+^8}{\ell^8}+{\cal O}\left(\frac{r_+^{10}}{\ell^{10}}\right)\right]-\epsilon^2\left[\frac{4135}{37128}\right.\\
&\left.+\frac{234006491065}{5686264584}\frac{r_+^6}{\ell^6}+{\cal O}\left(\frac{r_+^8}{\ell^8}\right)\right]-\epsilon^4\left[\frac{2720452858771243}{47290797450344448}+{\cal O}\left(\frac{r_+^6}{\ell^6}\right)\right]+{\cal O}(\epsilon^6)\nonumber\\
T_{H;7}={}&\frac{1}{\pi r_+}\left(\left[\frac32-\frac{397}{4}\frac{r_+^2}{\ell^2}-\frac{33003}{16}\frac{r_+^4}{\ell^4}+{\cal O}\left(\frac{r_+^6}{\ell^6}\right)\right]-\epsilon^2\left[\frac{3}{14}+\left\{\frac{114449}{12376}-\frac{3K^g_{7;2,2}}4+\frac{\pi\Gamma^2\big[\frac{7}{6}\big]}{2^{5/3}\Gamma^2\big[\frac{2}{3}\big]}\right\}\frac{r_+^2}{\ell^2}\right.\right.\nonumber\\
&\left.\left.+{\cal O}\left(\frac{r_+^4}{\ell^4}\right)\right]-\epsilon^4\left[\frac{1426337050717}{9906736519680}+{\cal O}\left(\frac{r_+^2}{\ell^2}\right)\right]\right)\\
S_7={}&r_+^7\frac{\pi^4}{12}\left(\left[1+\frac{81}{2}\frac{r_+^2}{\ell^2}+\frac{16839}{8}\frac{r_+^4}{\ell^4}+{\cal O}\left(\frac{r_+^6}{\ell^6}\right)\right]+\epsilon^2\left[\frac{75111}{12376}\frac{r_+^2}{\ell^2}+{\cal O}\left(\frac{r_+^4}{\ell^4}\right)\right]+\epsilon^4{\cal O}\left(\frac{r_+^2}{\ell^2}\right)+{\cal O}(\epsilon^6)\right)
\end{align}

\subsection*{n=9}

\begin{align}
E_9={}&\frac{\pi^4\ell^8}{32}\left(\left[\frac{3}{2}\frac{r_+^8}{\ell^8}+183\frac{r_+^{10}}{\ell^{10}}+{\cal O}\left(\frac{r_+^{12}}{\ell^{12}}\right)\right]+\epsilon^2\left[\frac{11}{756}+\frac{413843}{84672}\frac{r_+^8}{\ell^8}+{\cal O}\left(\frac{r_+^{10}}{\ell^{10}}\right)\right]+\epsilon^4\left[\frac{1065755263141}{651135350436480}\right.\right.\\
&\left.+{\cal O}\left(\frac{r_+^8}{\ell^8}\right)\right]+{\cal O}(\epsilon^6)\bigg)\nonumber\\
J_9={}&\frac{\pi^4\ell^9}{96}\left(\left[55\frac{r_+^{10}}{\ell^{10}}+{\cal O}\left(\frac{r_+^{12}}{\ell^{12}}\right)\right]+\epsilon^2\left[\frac1{252}+\frac{38905}{28224}\frac{r_+^8}{\ell^8}+{\cal O}\left(\frac{r_+^{10}}{\ell^{10}}\right)\right]+\epsilon^4\left[\frac{1088094883717}{2387496284933760}+{\cal O}\left(\frac{r_+^8}{\ell^8}\right)\right]\right.\nonumber\\
&+{\cal O}(\epsilon^6)\bigg)
\end{align}

\begin{align}
\ell\omega_9={}&\left[11-126\frac{r_+^8}{\ell^8}-\left\{12600+\frac{2520\Gamma\big[\frac{-5}{4}\big]\Gamma^2\big[\frac98\big]}{\Gamma\big[\frac{5}{4}\big]\Gamma^2\big[\frac{-1}8\big]}\right\}\frac{r_+^{10}}{\ell^{10}}+{\cal O}\left(\frac{r_+^{12}}{\ell^{12}}\right)\right]-\epsilon^2\left[\frac{754}{14535}\right.\\
&\left.+\frac{11400812957467}{99518819400}\frac{r_+^8}{\ell^8}+{\cal O}\left(\frac{r_+^{10}}{\ell^{10}}\right)\right]-\epsilon^4\left[\frac{971718029741243000453}{58241769402253879260000}+{\cal O}\left(\frac{r_+^8}{\ell^8}\right)\right]+{\cal O}(\epsilon^6)\nonumber\\
T_{H;9}={}&\frac{1}{\pi r_+}\left(\left[2-179\frac{r_+^2}{\ell^2}-\frac{12529}{4}\frac{r_+^4}{\ell^4}+{\cal O}\left(\frac{r_+^6}{\ell^6}\right)\right]-\epsilon^2\left[\frac{2}{9}+\left\{\frac{417847}{29070}-K^g_{9;2,2}+\frac{\pi\Gamma^2\big[\frac{9}{8}\big]}{2^{3/2}\Gamma^2\big[\frac{5}{8}\big]}\right\}\frac{r_+^2}{\ell^2}\right.\right.\nonumber\\
&\left.\left.+{\cal O}\left(\frac{r_+^4}{\ell^4}\right)\right]-\epsilon^4\left[\frac{1484561347453}{13416250352400}+{\cal O}\left(\frac{r_+^2}{\ell^2}\right)\right]\right)\\
S_9={}&r_+^9\frac{\pi^5}{48}\left(\left[1+\frac{121}{2}\frac{r_+^2}{\ell^2}+\frac{32351}{8}\frac{r_+^4}{\ell^4}+{\cal O}\left(\frac{r_+^6}{\ell^6}\right)\right]+\epsilon^2\left[\frac{214357}{29070}\frac{r_+^2}{\ell^2}+{\cal O}\left(\frac{r_+^4}{\ell^4}\right)\right]+\epsilon^4{\cal O}\left(\frac{r_+^2}{\ell^2}\right)+{\cal O}(\epsilon^6)\right)
\end{align}

\section{Perturbative Fields}

In this Appendix we catalogue all of the gravitational and scalar fields for the perturbative boson stars and hairy black holes in spacetime dimension $D=n+2$ for $n=3,5,7,9$.  The fields are labeled as $F_{n;p,q}$, where $p$ denotes the order in $\epsilon$ and $q$ denotes the order in $r_+$.  Note that $\mathrm{Li}_2[x]$ is the dilogarithm function while $P_\nu[x]$ is the Legendre function of the first kind.  

\subsubsection*{n=3 results}

\begin{align}
\tilde{s}_3(x)={}&22x^6+31x^4-250x^2+75\nonumber\\
\Pi^{out}_{3;1,2}={}&\frac14 \left(-\ell^2 (\ell^2 + 11 r^2) + 21 r^2 (\ell^2 + r^2) \log\left[1 + \frac{\ell^2}{r^2}\right]\right)\nonumber\\
\Pi^{out}_{3;1,4}={}&\frac{-\ell^2}{64(r^2+\ell^2)}\bigg(2r^{10}+(19+4\pi^2)r^8\ell^2+2(2163+4\pi^2r^6\ell^4+4(1156+\pi^2)r^4\ell^6+600r^2\ell^8+5\ell^{10}\bigg)\nonumber\\&
-\frac{1}{32}\bigg(24r^4\ell^4(r^2+\ell^2)\log^2\left[\frac{r}{\ell}\right]+2\ell^4(r^2+\ell^2)\log\left[\frac{r_+}{4\ell}\right]\left(8r^2\ell^2+\ell^4+12r^4\log\left[1+\frac{\ell^2}{r^2}\right]\right)\nonumber\\&
-\log\left[1+\frac{\ell^2}{r^2}\right]\left(r^{10}+9r^8\ell^2+3745r^6\ell^4+3267r^4\ell^6-51r^2\ell^8-\ell^{10}+441r^4\ell^4(r^2+\ell^2)\log\left[1+\frac{\ell^2}{r^2}\right]\right)\bigg)\nonumber\\&
-\frac{3r^4\ell^4}{8}(r^2+\ell^2)\mathrm{Li}_2\left[-\frac{r^2}{\ell^2}\right]\nonumber\\
f_{3;2,0}={}&\frac{5r^4+20r^2\ell^2+6\ell^4}{9}\nonumber\\
\Omega_{3;2,0}={}&\frac{r^4+4r^2\ell^2+6\ell^4}{12}\nonumber\\
f^{in}_{3;2,2}={}&\frac{2120+21\pi^2x^4P_{\frac12}\big[2x^2-1\big]\left(2\big(1-2x^2\big)P_{\frac12}\big[2x^2-1\big]+3P_{\frac32}\big[2x^2-1\big]\right)}{168x^3}\nonumber\\
g^{in}_{3;2,2}={}&-\frac{145}{14}\left(1-\frac{\ell^4}{z^4}\right)\nonumber\\
\Omega^{in}_{3;2,2}={}&\frac{190}{21}+\frac{25y^2}{7}\nonumber\\
f^{out}_{3;2,2}={}&-\frac{7}{6}r^2(r^2 + \ell^2)(5r^4+20r^2\ell^2+6\ell^4)\log\left[1 + \frac{\ell^2}{r^2}\right] - \frac{1}{72} \bigg(191r^{8} + 535r^6\ell^2 +20r^4\ell^4-84r^2\ell^6-48\ell^8\bigg)\nonumber\\
g^{out}_{3;2,2}={}&(r^2 + \ell^2)\left(\frac{20}{3}(r^2+\ell^2)^5 - 7 \ell^8 (4 r^2 + \ell^2)\right)\log\left[1 + \frac{\ell^2}{r^2}\right] - \frac{\ell^2}{9} \bigg(60 r^{10} + 330r^8\ell^2 +740r^6\ell^4+855 r^4\ell^6\nonumber\\
&+378r^2\ell^8+83\ell^{10}\bigg)\nonumber\\
\Omega^{out}_{3;2,2}={}&(r^2 + \ell^2)\left(-4(r^2+\ell^2)^4 +\frac{7\ell^4}{8} (r^4+4r^2\ell^2+6\ell^4)\right)\log\left[1+\frac{\ell^2}{r^2}\right]+\frac{\ell^2}{96}\bigg(384r^{8} + 1771r^6\ell^2\nonumber\\
&+3139r^4\ell^4+2516r^2\ell^6+668\ell^{8}\bigg)\nonumber\\
\Pi_{3;3,0}={}&\frac{900r^6+3935r^4\ell^2+5548r^2\ell^4+1540\ell^6}{2016}\nonumber\\
\Pi^{out}_{3;3,2}={}&\frac{-\ell^2}{48384}\bigg(14196r^{14}+7098(4\pi^2+25)r^{12}\ell^2+14(10140\pi^2-6037)r^{10}\ell^4+1183(240\pi^2-1667)r^{8}\ell^6\nonumber\\&
+65(4368\pi^2-55327)r^{6}\ell^8+5(28392 \pi^2-450949)r^{4}\ell^{10}+3(9464\pi^2-180149)r^{2}\ell^{12}+9240\ell^{14}\bigg)\nonumber\\&
+\frac{r^3}{8064}(r^2 + \ell^2)\log\left[1 + \frac{\ell^2}{r^2}\right] \bigg(2366r^{12} + 28392r^{10}\ell^2+ 34096r^8\ell^4 - 43516r^6\ell^6 -8889r^4\ell^8\nonumber\\&
 +145204r^2\ell^{10}+41208\ell^{12}\bigg)-\frac{169}{96}r^2\ell^4(r^2 + \ell^2)^5\left(2\mathrm{Li}_2\left[-\frac{r^2}{\ell^2}\right]-\log\left[1+\frac{r^2}{\ell^2}\right]\log\left[\frac{\ell^2(r^2 + \ell^2)}{r^4}\right]\right)\nonumber\\
 f_{3;4,0}={}&\frac1{1270080}\bigg(514952r^{14}+4631027r^{12}\ell^2+18512283r^{10}\ell^4+40913902r^8\ell^6+51954798r^6\ell^8\nonumber\\
&+36154839r^4\ell^{10}+11249595r^2\ell^{12}+1315860\ell^{14}\bigg)\nonumber\\
g_{3;4,0}={}&\frac1{1270080}\bigg(3541r^{16}+35410r^{14}\ell^2+149055r^{12}\ell^4+3052440r^{10}\ell^6+16099475r^8\ell^8+34403186r^6\ell^{10}\nonumber\\&+25939155r^4\ell^{12}+7971520r^2\ell^{14}+872290\ell^{16}\bigg)\nonumber\\
h_{3;4,0}={}&\frac{3541r^{12}+31869r^{10}\ell^2+123066r^8\ell^4+260694r^6\ell^6+311661r^4\ell^8+183645r^2\ell^{10}+22260\ell^{12}}{1270080}\nonumber\\
\Omega_{3;4,0}={}&\frac1{2540160}\bigg(167242r^{14}+1505178r^{12}\ell^2+6020712r^{10}\ell^4+14139048r^8\ell^6+20982192r^6\ell^8\nonumber\\&+19004760r^4\ell^{10}+8795055r^2\ell^{12}+1598455\ell^{14}\bigg)\nonumber\\
\Pi_{3;5,0}={}&\frac1{853493760}\bigg(428716940r^{18}+4416801537r^{16}\ell^2+20395866890r^{14}\ell^4+55586393870r^{12}\ell^6\nonumber\\&+98320298706r^{10}\ell^8+115794392980r^8\ell^{10}+88872056182r^6\ell^{12}+41756607180r^4\ell^{14}\nonumber\\
&+10678880150r^2\ell^{16}+1128452101\ell^{18}\bigg)\nonumber
\end{align}

\subsubsection*{n=5 results}

\begin{align}
\tilde{s}_5(x)={}&44x^{12}+88x^{10}+113x^8-1210x^6-784x^4+490x^2+245\nonumber\\
\Pi^{out}_{5;1,4}={}&\frac{1}{8}\left(\ell^2\big(100r^4+71r^2\ell^2-\ell^4\big)- 30 r^4 (\ell^2 + r^2) \log\left[1 + \frac{\ell^2}{r^2}\right]\right)\nonumber\\
\Pi^{out}_{5;1,6}={}&\frac{\ell^{8}}{6}\bigg(3456r^4+2295r^2\ell^2-111\ell^4+(r^2+\ell^2)(45r^4+9r^2\ell^2+\ell^4)\frac{\Gamma^2\left[\frac{1}{4}\right]}{\Gamma^2\left[\frac{-1}{4}\right]}\bigg)\nonumber\\&
-r^6\ell^6(r^2+\ell^2)\left(261-10\frac{\Gamma^2\left[\frac{1}{4}\right]}{\Gamma^2\left[\frac{-1}{4}\right]}\right)\log\left[1+\frac{\ell^2}{r^2}\right]\nonumber\\
f_{5;2,0}={}&\frac{7r^6+42r^4\ell^2+105r^2\ell^4+20\ell^6}{50}\nonumber\\
\Omega_{5;2,0}={}&\frac{r^6+6r^4\ell^2+15r^2\ell^4+20\ell^6}{60}\nonumber\\
f^{in}_{5;2,2}={}&\frac{19831\pi(1+x^4)+8580\Gamma^4\big[\frac{5}{4}\big]x^6P_{\frac14}\big[2x^4-1\big]\left(3\big(1-2x^4\big)P_{\frac14}\big[2x^4-1\big]+5P_{\frac54}\big[2x^4-1\big]\right)}{2145\pi x^3}\nonumber\\
g^{in}_{5;2,2}={}&-\int_1^{z/\ell}{\frac{14\big(2458 + 4916 x^2 + 3173 x^4\big)}{715x^7(1+x^2)^2}dx}\nonumber\\
\Omega^{in}_{5;2,2}={}&\frac{67368}{715}-\frac{47537y^4}{715}\nonumber\\
f^{out}_{5;2,4}={}&\frac{3}{20}r^4(r^2 + \ell^2)(7r^6+42r^4\ell^2+105r^2\ell^4+20\ell^6)\log\left[1 + \frac{\ell^2}{r^2}\right] - \frac{1}{1200} \bigg(3463r^{12}+25501r^{10}\ell^2\nonumber\\&
+80913r^8\ell^4+143675r^6\ell^6+91100r^4\ell^8+13140r^2\ell^{10}-360\ell^{12}\bigg)\nonumber\\
g^{out}_{5;2,4}={}&-r^2(r^2 + \ell^2)\left(\frac{133}{5}(r^2+\ell^2)^7 - 3 \ell^{12} (6 r^2 + \ell^2)\right) \log\left[1 + \frac{\ell^2}{r^2}\right] +\frac{\ell^2}{300} \bigg(7980 r^{16} + 59850r^{14}\ell^2 \nonumber\\
&+194180r^{12}\ell^4+354445 r^{10}\ell^6+395276r^8\ell^8+272118r^6\ell^{10}+92268r^4\ell^{12}+7401r^2\ell^{14}-90\ell^{16}\bigg)\nonumber\\
\Omega^{out}_{5;2,4}={}&(r^2 + \ell^2)\left(\frac{169}{6}(r^2+\ell^2)^6 -\frac{\ell^6}{8} (r^6+6r^4\ell^2+15r^2\ell^4+20\ell^6)\right)\log\left[1+\frac{\ell^2}{r^2}\right]-\frac{\ell^2}{1440}\bigg(40560r^{14}\nonumber\\
&+263640r^{12}\ell^2+722791r^{10}\ell^4+1074337r^{8}\ell^6+918573r^6\ell^{8}+430519r^4\ell^{10}+81352r^2\ell^{12}-3780\ell^{14}\bigg)\nonumber\\
\Pi_{5;3,0}={}&\frac{53970r^{10}+313026r^8\ell^2+773598r^6\ell^4+1035891r^4\ell^6+767826r^2\ell^8+152581\ell^{10}}{257400}\nonumber\\
\Pi^{out}_{5;3,4}={}&\frac{\ell^2}{883396800}\bigg(464785464r^{22}+7204174692r^{20}\ell^2+309856976(15\pi^2+182)r^{18}\ell^4\nonumber\\&
+2(16267491240\pi^2+99389066437)r^{16}\ell^6+2(48802473720\pi^2+175884179161)r^{14}\ell^8\nonumber\\&
+88(1848578550\pi^2+3555421769)r^{12}\ell^{10}+3080(52816530\pi^2+36628159)r^{10}\ell^{12}\nonumber\\&
+4(24401236860\pi^2+2936476817)r^{8}\ell^{14}+(32534982480\pi^2+44985717097)r^{6}\ell^{16}\nonumber\\&
+429(10834160 \pi^2+89900771)r^{4}\ell^{18}+5998786365r^{2}\ell^{20}-65457249\ell^{22}\bigg)\nonumber\\&
-\frac{r^4}{1029600}(r^2 + \ell^2)\log\left[1 + \frac{\ell^2}{r^2}\right] \bigg(541708r^{18}+8125620r^{16}\ell^2+61754712r^{14}\ell^4+221505790r^{12}\ell^6\nonumber\\&
+ 421393950r^{10}\ell^8+447202548r^8\ell^{10} +255856550r^6\ell^{12} +66949977r^4\ell^{14}+6608370r^2\ell^{16}\nonumber\\&
-1853965\ell^{18}\bigg)+\frac{135427}{8580}r^4\ell^6(r^2 + \ell^2)^7\left(2\mathrm{Li}_2\left[-\frac{r^2}{\ell^2}\right]-\log\left[1+\frac{r^2}{\ell^2}\right]\log\left[\frac{\ell^2(r^2 + \ell^2)}{r^4}\right]\right)\nonumber\\
f_{5;4,0}={}&\frac1{618377760000}\bigg(35065241462r^{20}+455760809811r^{18}\ell^2+2733953554501r^{16}\ell^4\nonumber\\
&+10021950443882r^{14}\ell^6+24529389790620r^{12}\ell^8 +41236810544215r^{10}\ell^{10}+47555951613885r^8\ell^{12}\nonumber\\
&+36498131558460r^6\ell^{14}+17103932986140r^4\ell^{16}+3682611572640r^2\ell^{18}+295557662400\ell^{20}\bigg)\nonumber\\
g_{5;4,0}={}&\frac1{618377760000}\bigg(87329195r^{22}+1222608730r^{20}\ell^2+7946956745r^{18}\ell^4+31598878220r^{16}\ell^6\nonumber\\&+677658846865r^{14}\ell^8+4244040931130r^{12}\ell^{10}+12564927910535r^{10}\ell^{12}+21539953150144r^8\ell^{14}\nonumber\\&+22932404913236r^6\ell^{16}+11564746522784r^4\ell^{18}+2586519623296r^2\ell^{20}+198795488864\ell^{22}\bigg)\nonumber\\
h_{5;4,0}={}&\frac1{123675552000}\bigg(17465839r^{18}+227055907r^{16}\ell^2+1362335442r^{14}\ell^4+4978052794r^{12}\ell^6\nonumber\\&+12277654675r^{10}\ell^8+21295887327r^8\ell^{10}+25938182556r^6\ell^{12}+20983798836r^4\ell^{14}\nonumber\\&
+9390693312r^2\ell^{16}+769728960\ell^{18}\bigg)\nonumber\\
\Omega_{5;4,0}={}&\frac1{46378332000}\bigg(327248543r^{20}+4254231059r^{18}\ell^2+25525386354r^{16}\ell^4+93593083298r^{14}\ell^6\nonumber\\&+234379619045r^{12}\ell^8+420702504651r^{10}\ell^{10}+547149789758r^8\ell^{12}+505325794688r^6\ell^{14}\nonumber\\&+311179210446r^4\ell^{16}+106176159440r^2\ell^{18}+15058082990\ell^{20}\bigg)\nonumber\\
\Pi_{5;5,0}={}&\frac1{453635740958400000}\bigg(74253000574956420r^{26}+1032668691070620996r^{24}\ell^2\nonumber\\&+6684077146010747418r^{22}\ell^4+26698941169899723207r^{20}\ell^6+73557029936994344181r^{18}\ell^8\nonumber\\&+147920037831416512914r^{16}\ell^{10}+223871564194330948248r^{14}\ell^{12}+258620332265962810395r^{12}\ell^{14}\nonumber\\&+228077768805842399885r^{10}\ell^{16}+151159034180233581556r^8\ell^{18}+72350175319616674844r^6\ell^{20}\nonumber\\&+23112600885726752792r^4\ell^{22}+4271077136958547132r^2\ell^{24}+337633104499816268\ell^{26}\bigg)\nonumber
\end{align}

\subsubsection*{n=7 results}

\begin{align}
\tilde{s}_7(x)={}&74x^{18}+148x^{16}+222x^{14}+271x^{12}-3860x^{10}-2807x^{8}-1782x^6+1701x^4+1134x^2+567\nonumber\\
\Pi^{out}_{7;1,6}={}&\frac{1}{24}\left(-\ell^2\big(42r^6-231r^4\ell^2-163r^2\ell^4+2\ell^6\big)+798 r^6 (\ell^2 + r^2) \log\left[1 + \frac{\ell^2}{r^2}\right]\right)\nonumber\\
\Pi^{out}_{7;1,8}={}&\frac{\ell^{10}}{3}\bigg(-1200r^8+1416r^6\ell^2+1448r^4\ell^4-61r^2\ell^6+\frac{\Gamma\left[\frac{-4}{3}\right]\Gamma^2\left[\frac{7}{6}\right]}{\Gamma\left[\frac{4}{3}\right]\Gamma^2\left[\frac{-1}{6}\right]}\big(672r^8+840r^6\ell^2+200r^4\ell^4\nonumber\\&
+35r^2\ell^6+3\ell^8\big)\bigg)+8r^8\ell^8(r^2+\ell^2)\left(35\frac{\Gamma\left[\frac{-4}{3}\right]\Gamma^2\left[\frac{7}{6}\right]}{\Gamma\left[\frac{4}{3}\right]\Gamma^2\left[\frac{-1}{6}\right]}+302\right)\log\left[1+\frac{\ell^2}{r^2}\right]\nonumber\\
f_{7;2,0}={}&\frac{9r^8+72r^6\ell^2+252r^4\ell^4+504r^2\ell^6+70\ell^8}{245}\nonumber\\
\Omega_{7;2,0}={}&\frac{r^8+8r^6\ell^2+28r^4\ell^4+56r^2\ell^6+70\ell^8}{280}\nonumber\\
f^{in}_{7;2,2}={}&\frac{75111\Gamma^2\big[\frac{4}{3}\big](1+2x^6)+6188\Gamma^4\big[\frac{7}{6}\big]x^8P_{\frac16}\big[2x^6-1\big]\left(4\big(1-2x^6\big)P_{\frac16}\big[2x^6-1\big]+7P_{\frac76}\big[2x^6-1\big]\right)}{9282\Gamma^2\big[\frac{4}{3}\big] x^3}\nonumber\\
g^{in}_{7;2,2}={}&-\int_1^{z/\ell}{\frac{6 \big(13417 + 26834 x^2 + 40251 x^4 + 45398 x^6 + 22699 x^8\big)}{1547 x^9 (1 + x^2 + x^4)^2}}\nonumber\\
\Omega^{in}_{7;2,2}={}&\frac{439884}{1547}-\frac{364773y^6}{1547}\nonumber\\
f^{out}_{7;2,6}={}&-\frac{19}{70}r^6(r^2 + \ell^2)(9r^8+72r^6\ell^2+252r^4\ell^4+504r^2\ell^6+70\ell^8)\log\left[1 + \frac{\ell^2}{r^2}\right] - \frac{1}{58800} \bigg(181653r^{16}\nonumber\\&
+1491237r^{14}\ell^2+5318568r^{12}\ell^4+10686312r^{10}\ell^6+13013028r^{8}\ell^8+10922100r^{6}\ell^{10}+4645200r^{4}\ell^{12}\nonumber\\&
+518000r^2\ell^{14}-11200\ell^{16}\bigg)\nonumber\\
g^{out}_{7;2,6}={}&r^4(r^2 + \ell^2)\left(\frac{102}{7}(r^2+\ell^2)^9 - 19 \ell^{16} (8 r^2 + \ell^2)\right) \log\left[1 + \frac{\ell^2}{r^2}\right] -\frac{\ell^2}{2940} \bigg(42840 r^{22} + 406980r^{20}\ell^2 \nonumber\\
&+1727880r^{18}\ell^4+4308990 r^{16}\ell^6+6970068r^{14}\ell^8+7607670r^{12}\ell^{10}+5632440r^{10}\ell^{12}+ 2754765r^{8}\ell^{14}\nonumber\\
&+791090r^{6}\ell^{16}+220733r^{4}\ell^{18}+82250r^{2}\ell^{20}+350\ell^{22}\bigg)\nonumber\\
\Omega^{out}_{7;2,6}={}&-(r^2 + \ell^2)\left(\frac{743}{4}(r^2+\ell^2)^8 -\frac{19\ell^8}{80} (r^8+8r^6\ell^2+28r^4\ell^4+56r^2\ell^6+70\ell^8)\right)\log\left[1+\frac{\ell^2}{r^2}\right]\nonumber\\
&+\frac{\ell^2}{67200}\bigg(12482400r^{20}+106100400r^{18}\ell^2+397356400r^{16}\ell^4+858186117r^{14}\ell^6+1172895573r^{12}\ell^8\nonumber\\&+1044569272r^{10}\ell^{10}+598459448r^{8}\ell^{12}+207032712r^6\ell^{14}+36525880r^4\ell^{16}+2024400r^2\ell^{18}\nonumber\\
&+126000\ell^{20}\bigg)\nonumber\\
\Pi_{7;3,0}={}&\frac1{62375040}\bigg(6252120r^{14}+48974940r^{12}\ell^2+168013320r^{10}\ell^4+331552839r^8\ell^6+415360056r^6\ell^8\nonumber\\&+342224260r^4\ell^{10}+181648328r^2\ell^{12}+28301260\ell^{14}\bigg)\nonumber\\
\Pi^{out}_{7;3,6}={}&\frac{\ell^2}{36392093337600}\bigg(8291208954600r^{30}+158914838296500r^{28}\ell^2+1479980798396100r^{26}\ell^4\nonumber\\&
+230311359850(1680\pi^2+39899)r^{24}\ell^6+30(116076925364400\pi^2+1084763808099529)r^{22}\ell^8\nonumber\\&
+420(33164835818400\pi^2+141178710556613)r^{20}\ell^{10}+20(1625076955101600\pi^2\nonumber\\&
+1492281977949773)r^{18}\ell^{12}+3(16250769551016000\pi^2-31895994916390159)r^{16}\ell^{14}\nonumber\\&
+45(1083384636734400\pi^2-5183425872436277)r^{14}\ell^{16}+6(5416923183672000\pi^2\nonumber\\&
-41728761232798273)r^{12}\ell^{18}+6(2321538507288000\pi^2-24834461884681829)r^{10}\ell^{20}\nonumber\\&
+9(386923084548000\pi^2-5065698248265511)r^{8}\ell^{22}+12155(31832421600\pi^2-242104978313)r^{6}\ell^{24}\nonumber\\&
+1676613343633800r^{4}\ell^{26}+151166509614120r^{2}\ell^{28}-1376007261200\ell^{30}\bigg)\nonumber\\&
-\frac{r^6}{249500160}(r^2 + \ell^2)\log\left[1 + \frac{\ell^2}{r^2}\right] \bigg(56843610r^{24}+1061080720r^{22}\ell^2+9625517960r^{20}\ell^4\nonumber\\&
+58359439600r^{18}\ell^6+202222134520r^{16}\ell^8+364268311400r^{14}\ell^{10}+204109535140r^{12}\ell^{12}\nonumber\\&
-475573634440r^{10}\ell^{14}-1194627217511r^8\ell^{16}-1278670997224r^6\ell^{18}-779269351420r^4\ell^{20}\nonumber\\&
-273392960792r^2\ell^{22}-37998233560\ell^{24}\bigg)+\frac{9473935}{297024}r^6\ell^8(r^2 + \ell^2)^9\left(2\mathrm{Li}_2\left[-\frac{r^2}{\ell^2}\right]\right.\nonumber\\&
\left.-\log\left[1+\frac{r^2}{\ell^2}\right]\log\left[\frac{\ell^2(r^2 + \ell^2)}{r^4}\right]\right)\nonumber\\
f_{7;4,0}={}&\frac1{371502619488000}\bigg(2968444009876r^{26}+50460227131621r^{24}\ell^2+403651927726529r^{22}\ell^4\nonumber\\
&+2018090265782824r^{20}\ell^6+7062641146585640r^{18}\ell^8+18211924445078266r^{16}\ell^{10}\nonumber\\
&+35302266172890626r^{14}\ell^{12}+51618427799679364r^{12}\ell^{14}+56539135502871164r^{10}\ell^{16}\nonumber\\&+45556296988624565r^8\ell^{18}+25965374820716045r^6\ell^{20}+9521410725967140r^4\ell^{22}\nonumber\\&
+1589105610832740r^2\ell^{24}+96780316472840\ell^{26}\bigg)\nonumber\\
g_{7;4,0}={}&\frac1{371502619488000}\bigg(3321036271r^{28}+59778652878r^{26}\ell^2+508118549463r^{24}\ell^4+2709965597136r^{22}\ell^6\nonumber\\&+10157293967310r^{20}\ell^8+193867029381396r^{18}\ell^{10}+1535551141626138r^{16}\ell^{12}+5953776409345128r^{14}\ell^{14}\nonumber\\&+13769626720091151r^{12}\ell^{16}+20860826548147390r^{10}\ell^{18}+21693866896535335r^8\ell^{20}\nonumber\\&+15807036948862760r^6\ell^{22}+6104455301649630r^4\ell^{24}+1080779423604460r^2\ell^{26}+63735166423850\ell^{28}\bigg)\nonumber\\
h_{7;4,0}={}&\frac1{53071802784000}\bigg(474433753r^{24}+8065373801r^{22}\ell^2+64522990408r^{20}\ell^4+322614952040r^{18}\ell^6\nonumber\\&+1128813864010r^{16}\ell^8+2930312879858r^{14}\ell^{10}+5829937982596r^{12}\ell^{12}+9029812070708r^{10}\ell^{14}\nonumber\\&+10884488013685r^8\ell^{16}+9976953706005r^6\ell^{18}+6500468085540r^4\ell^{20}+2477855239860r^2\ell^{22}\nonumber\\&+153269396280\ell^{24}\bigg)\nonumber\\
\Omega_{7;4,0}={}&\frac1{25474465336320}\bigg(20377852226r^{26}+346423487842r^{24}\ell^2+2771387902736r^{22}\ell^4+13856939513680r^{20}\ell^6\nonumber\\&+48499288297880r^{18}\ell^8+126161196878808r^{16}\ell^{10}+252029223792528r^{14}\ell^{12}+391916979706512r^{12}\ell^{14}\nonumber\\&+473620370087280r^{10}\ell^{16}+438282129396128r^8\ell^{18}+301265955958360r^6\ell^{20}+144325351642808r^4\ell^{22}\nonumber\\&
+39868495607983r^2\ell^{24}+4684576834175\ell^{26}\bigg)\nonumber\\
\Pi_{7;5,0}={}&\frac1{21848348422059134976000}\bigg(1117614384817280519400r^{34}+20022251781128501589300r^{32}\ell^2\nonumber\\&+169475537803320197519400r^{30}\ell^4+900812314937847677993001r^{28}\ell^6\nonumber\\&+3370660892553798052350570r^{26}\ell^8+9433307528652764598206937r^{24}
\ell^{10}\nonumber\\&+20477493617041202568028488r^{22}\ell^{12}+35284523399336392108933005r^{20}\ell^{14}\nonumber\\&+48974717480356234743070466r^{18}\ell^{16}+55227488952051138771625005r^{16}\ell^{18}\nonumber\\&
+50757619733081920144022472r^{14}\ell^{20}+37918308373546561272000663r^{12}\ell^{22}\nonumber\\&+22792336247876448363452550r^{10}\ell^{24}+10789673943068638679272455r^8\ell^{26}\nonumber\\&+3856281625419549220592040r^6\ell^{28}+957479816306705256586845r^4\ell^{30}\nonumber\\&
+141066092907373532097690r^2\ell^{32}+8982162145045598581865\ell^{34}\bigg)\nonumber
\end{align}

\subsubsection*{n=9 results}

\begin{align}
\tilde{s}_9(x)={}&112x^{24}+224x^{22}+336x^{20}+448x^{18}+529x^{16}-9514x^{14}-7457x^{12}-5400x^{10}-3388x^8\nonumber\\
&+4356x^6+3267x^4+2178x^2+1089\nonumber\\
\Pi^{out}_{9;1,8}={}&\frac{1}{16}\left(\ell^2\big(1236r^{8}+1278r^6\ell^2+234r^4\ell^4+103r^2\ell^6-\ell^8\big)+612 r^8 (\ell^2 + r^2) \log\left[1 + \frac{\ell^2}{r^2}\right]\right)\nonumber\\
\Pi^{out}_{9;1,10}={}&\frac{\ell^{12}}{12}\bigg(99300r^{10}+99150r^8\ell^2+16450r^6\ell^4+8275r^4\ell^6-273r^2\ell^8+\frac{\Gamma\left[\frac{-5}{4}\right]\Gamma^2\left[\frac{9}{8}\right]}{\Gamma\left[\frac{5}{4}\right]\Gamma^2\left[\frac{-1}{8}\right]}\big(12600r^{10}\nonumber\\&
+16200r^8\ell^2+4500r^6\ell^4+1050r^4\ell^6+162r^2\ell^8+12\ell^{10}\big)\bigg)\nonumber\\&
+5r^{10}\ell^{10}(r^2+\ell^2)\left(252\frac{\Gamma\left[\frac{-5}{4}\right]\Gamma^2\left[\frac{9}{8}\right]}{\Gamma\left[\frac{5}{4}\right]\Gamma^2\left[\frac{-1}{8}\right]}+655\right)\log\left[1+\frac{\ell^2}{r^2}\right]\nonumber\\
f_{9;2,0}={}&\frac{11r^{10}+110r^8\ell^2+495r^6\ell^4+1320r^4\ell^6+2310r^2\ell^8+252\ell^{10}}{1134}\nonumber\\
\Omega_{9;2,0}={}&\frac{r^{10}+10r^8\ell^2+45r^6\ell^4+120r^4\ell^6+210r^2\ell^8+252\ell^{10}}{1260}\nonumber\\
f^{in}_{9;2,2}={}&\frac{214357\Gamma^2\big[\frac{5}{4}\big](1+3x^8)+14535\Gamma^4\big[\frac{9}{8}\big]x^{10}P_{\frac18}\big[2x^8-1\big]\left(5\big(1-2x^8\big)P_{\frac18}\big[2x^8-1\big]+9P_{\frac98}\big[2x^8-1\big]\right)}{29070\Gamma^2\big[\frac{5}{4}\big] x^3}\nonumber\\
g^{in}_{9;2,2}={}&-\int_1^{z/\ell}{\frac{22 \big(7322 + 11519 x^8 - 34884 x^{10} +16043 x^{16}\big)}{2907 x^{11} (x^8-1)^2}}\nonumber\\
\Omega^{in}_{9;2,2}={}&\frac{1740860}{2907}-\frac{1526503y^8}{2907}\nonumber\\
f^{out}_{9;2,8}={}&-\frac{17}{252}r^8(r^2 + \ell^2)(11r^{10}+110r^8\ell^2+495r^6\ell^4+1320r^4\ell^6+2310r^2\ell^8+252\ell^{10})\log\left[1 + \frac{\ell^2}{r^2}\right]\nonumber\\& - \frac{1}{635040} \bigg(2069215r^{20}
+22290125r^{18}\ell^2+108858805r^{16}\ell^4+317937015r^{14}\ell^6+616435380r^{12}\ell^8\nonumber\\&
+831907692r^{10}\ell^{10}+610190532r^{8}\ell^{12}+229099500r^{6}\ell^{14}+63018900r^{4}\ell^{16}+5600700r^{2}\ell^{18}-88200\ell^{20}\bigg)\nonumber\\
g^{out}_{9;2,8}={}&r^6(r^2 + \ell^2)\left(\frac{8107}{9}(r^2+\ell^2)^{11} - 17 \ell^{20} (10 r^2 + \ell^2)\right) \log\left[1 + \frac{\ell^2}{r^2}\right] -\frac{\ell^2}{22680}\bigg(20429640r^{28}\nonumber\\&+234940860r^{26}\ell^{2}+1232588280r^{24}\ell^{4}+3896953830r^{22}\ell^{6}+8257660488r^{20}\ell^{8}+12330649716r^{18}\ell^{10}\nonumber\\&
+13265840808r^{16}\ell^{12}+10313684865r^{14}\ell^{14}+5707490140r^{12}\ell^{16}+2161699122r^{10}\ell^{18}\nonumber\\&
+523189164r^{8}\ell^{20}+71641657r^{6}\ell^{22}+3249540r^{4}\ell^{24}+380520r^{2}\ell^{26}+1470\ell^{28}\bigg)\nonumber\\
\Omega^{out}_{9;2,8}={}&(r^2 + \ell^2)\left(\frac{12013}{10}(r^2+\ell^2)^{10} -\frac{17\ell^{10}}{280} (r^{10}+10r^8\ell^2+45r^6\ell^4+120r^4\ell^6+210r^2\ell^8+252\ell^{10})\right)\times\nonumber\\&
\times\log\left[1+\frac{\ell^2}{r^2}\right]-\frac{\ell^2}{705600}\bigg(847637280r^{26}+8900191440r^{24}\ell^2+42240591120r^{22}\ell^4\nonumber\\&
+119446220040r^{20}\ell^6+223168574011r^{18}\ell^8+288434742161r^{16}\ell^{10}+261960125065r^{14}\ell^{12}\nonumber\\&
+165920263095r^{12}\ell^{14}+70798602040r^{10}\ell^{16}+18754765232r^{8}\ell^{18}+2483120752r^6\ell^{20}\nonumber\\&
+29914500r^4\ell^{22}-14391300r^2\ell^{24}-1131900\ell^{26}\bigg)\nonumber\\
\Pi_{9;3,0}={}&\frac1{1025589600}\bigg(48768720r^{18}+480720240r^{16}\ell^2+2135373240r^{14}\ell^4+5636851220r^{12}\ell^6\nonumber\\&+9822117536r^{10}\ell^8+11878432160r^8\ell^{10}+10221019320r^6\ell^{12}+6313029495r^4\ell^{14}\nonumber\\&
+2750576510r^2\ell^{16}+356786123\ell^{18}\bigg)\nonumber\\
\Pi^{out}_{9;3,8}={}&\frac{-\ell^2}{425578660416000}\bigg(1038886376760960r^{38}+23894386665502080r^{36}\ell^2+266041486315535840r^{34}\ell^4\nonumber\\&
+1923324978843457280r^{32}\ell^6+8657386473008(25200\pi^2+1195469)r^{30}\ell^8\nonumber\\&
+56(42854063041389600\pi^2+723954851343998297)r^{28}\ell^{10}+87780(136695575889600\pi^2\nonumber\\&
+1267804484857559)r^{26}\ell^{12}+1260( 28569375360926400\pi^2+166649848543617091)r^{24}\ell^{14}\nonumber\\&
+210(342832504331116800\pi^2+1261851310575329737)r^{22}\ell^{16}+42(2399827530317817600\pi^2\nonumber\\&
+4862510402105676583)r^{20}\ell^{18}+4158(24240682124422400\pi^2+13837058565923907)r^{18}\ell^{20}\nonumber\\&
+10(7199482590953452800\pi^2-6011370596731037917)r^{16}\ell^{22}+10(3599741295476726400\pi^2\nonumber\\&
-8283514692990005221)r^{14}\ell^{24}+190(63153356060995200\pi^2-250880065213913363)r^{12}\ell^{26}\nonumber\\&
+128(18748652580607950\pi^2-116667459449721109)r^{10}\ell^{28}+71136(3066887920600\pi^2\nonumber\\&
-33694272545267)r^{8}\ell^{30}-163111679979258465r^{6}\ell^{32}-18207747905205495r^{4}\ell^{34}\nonumber\\&
-1346064504085665r^{2}\ell^{36}+9253248100005\ell^{38}\bigg)-\frac{2629813}{1710}r^8\ell^{10}(r^2 + \ell^2)^{11}\left(2\mathrm{Li}_2\left[-\frac{r^2}{\ell^2}\right]\right.\nonumber\\&
\left.-\log\left[1+\frac{r^2}{\ell^2}\right]\log\left[\frac{\ell^2(r^2 + \ell^2)}{r^4}\right]\right)+\frac{r^8}{4102358400}(r^2 + \ell^2)\log\left[1 + \frac{\ell^2}{r^2}\right]\bigg(10014327904r^{30}\nonumber\\&
+225322377840r^{28}\ell^2+2453510336480r^{26}\ell^4+17349823093680r^{24}\ell^6+91480885403040r^{22}\ell^8\nonumber\\&
+360803126340498r^{20}\ell^{10}+1040279830382820r^{18}\ell^{12}+2178047537719290r^{16}\ell^{14}\nonumber\\&
+3321710525234040r^{14}\ell^{16}+ 3698150956946320r^{12}\ell^{18}+2991052821746808r^{10}\ell^{20}\nonumber\\&
+1729094535021140r^8\ell^{22}+690952595079480r^6\ell^{24}+179411207692455r^4\ell^{26}\nonumber\\&
+26968400204190r^2\ell^{28}+1682848959147\ell^{30}\bigg)\nonumber\\
f_{9;4,0}={}&\frac1{859498662576153600}\bigg(937254465799230r^{32}+19681794632595465r^{30}\ell^2+196811905684882635r^{28}\ell^4\nonumber\\&+1246433118183419250r^{26}\ell^6+5608737609387866100r^{24}\ell^8+19068904767424154265r^{22}\ell^{10}\nonumber\\&+50684524900929500343r^{20}\ell^{12}+107062480455455025288r^{18}\ell^{14}+180579919814518993740r^{16}\ell^{16}\nonumber\\&+242609187466772524790r^{14}\ell^{18}+257756682509308369842r^{12}\ell^{20}+213952704807502670100r^{10}\ell^{22}\nonumber\\&+135934002405968654888r^8\ell^{24}+63513374918908684770r^6\ell^{26}+19770363913882326030r^4\ell^{28}\nonumber\\&
+2720901281792621928r^2\ell^{30}+133340623834666248\ell^{32}\bigg)\nonumber\\
g_{9;4,0}={}&\frac1{859498662576153600}\bigg(549149188365r^{34}+12081282144030r^{32}\ell^2+126853462512315r^{30}\ell^4\nonumber\\&+845689750082100r^{28}\ell^6+4017026312889975r^{26}\ell^8+14460723267227622r^{24}\ell^{10}\nonumber\\&+219306855955192929r^{22}\ell^{12}+2041707149885182032r^{20}\ell^{14}+9866747321987487030r^{18}\ell^{16}\nonumber\\&+29286912971461691700r^16\ell^{18}+58562593998521246634r^{14}\ell^{20}+82904154261107854584r^{12}\ell^{22}\nonumber\\&+85422451772928300030r^{10}\ell^{24}+65138644777217991980r^8\ell^{26}+37242538451355126890r^6\ell^{28}\nonumber\\&+11936492827687331904r^4\ell^{30}+1768012312961670776r^2\ell^{32}+84495843563523392\ell^{34}\bigg)\nonumber\\
h_{9;4,0}={}&\frac1{95499851397350400}\bigg(61016576485r^{30}+1281348106185r^{28}\ell^2+12813481061850r^{26}\ell^4\nonumber\\&+81152046725050r^{24}\ell^6+365184210262725r^{22}\ell^8+1241596238094513r^{20}\ell^{10}\nonumber\\&+3310396957607208r^{18}\ell^{12}+7089196246488360r^{16}\ell^{14}+12381092092392030r^{14}\ell^{16}\nonumber\\&+17783794333160310r^{12}\ell^{18}+21034536810890148r^{10}\ell^{20}+20300574058072308r^8\ell^{22}\nonumber\\&+15541844723695290r^6\ell^{24}+8797952237444370r^4\ell^{26}+3015895473831240r^2\ell^{28}\nonumber\\&
+149725307211480\ell^{30}\bigg)\nonumber\\
\Omega_{9;4,0}={}&\frac1{59687407123344000}\bigg(5440474418585r^{32}+114249962790285r^{30}\ell^2+1142499627902850r^{28}\ell^4\nonumber\\&+7235830976718050r^{26}\ell^6+32561239395231225r^{24}\ell^8+110708213943786165r^{22}\ell^{10}\nonumber\\&+295268817053306040r^{20}\ell^{12}+632417867298394200r^{18}\ell^{14}+1101647489722639650r^{16}\ell^{16}\nonumber\\&+1565834578187766250r^{14}\ell^{18}+1807245829726785084r^{12}\ell^{20}+1674173405090497584r^{10}\ell^{22}\nonumber\\&+1221884704722504550r^8\ell^{24}+681865813132481250r^6\ell^{26}+273777806813998500r^4\ell^{28}\nonumber\\&+64940197030107672r^2\ell^{30}+6598585567818972\ell^{32}\bigg)\nonumber\\
\Pi_{9;5,0}={}&\frac1{12928610477426464084962297600000}\bigg(196278755076119819922240261840r^{42}\nonumber\\&+4302621620873952400266415533360r^{40}\ell^2+45035990011802132234460421933080r^{38}\ell^4\nonumber\\&+299476749663332693852942194632980r^{36}\ell^6+1419986713892478010760754706555122r^{34}\ell^8\nonumber\\&+5107957999464578930096857586903749r^{32}\ell^{10}+14480110437447756571401543030317327r^{30}\ell^{12}\nonumber\\&+33171546764185616749586054262623220r^{28}\ell^{14}+62477592200815789501550742105430020r^{26}\ell^{16}\nonumber\\&+97921202768932011707091446069275799r^{24}\ell^{18}+128775176906725704442407617007965163r^{22}\ell^{20}\nonumber\\&+142873004760249240821838450887004074r^{20}\ell^{22}+134129087251912145628690383503063720r^{18}\ell^{24}\nonumber\\&+106606113471306396324075091246911420r^{16}\ell^{26}+71564603098088232189439094493589402r^{14}\ell^{28}\nonumber\\&+40318687255770901091487447290055224r^{12}\ell^{30}+18828704309439914264745324330571272r^{10}\ell^{32}\nonumber\\&+7126896792786249923435515161591030r^8\ell^{34}+2095459575089784573411052833767630r^6\ell^{36}\nonumber\\&+438680016278348277072828543787740r^4\ell^{38}+54960718928326817581432660068800r^2\ell^{40}\nonumber\\&+2963660262582405195600812804780\ell^{42}\bigg)\nonumber
\end{align}

\bibliographystyle{plain}

\begin{thebibliography}{99}

\bibitem{Ridgway:1995ke}
  S.~A.~Ridgway and E.~J.~Weinberg,
  Phys.\ Rev.\  D {\bf 52}, 3440 (1995)
  [arXiv:gr-qc/9503035].

\bibitem{Hawking:1971vc}
  S.~W.~Hawking,
  Commun.\ Math.\ Phys.\  {\bf 25}, 152-166 (1972).
  
\bibitem{Hollands:2006rj}
  S.~Hollands, A.~Ishibashi, R.~M.~Wald,
  Commun.\ Math.\ Phys.\  {\bf 271}, 699-722 (2007).
  [gr-qc/0605106].
  
\bibitem{Moncrief:2008mr}
  V.~Moncrief, J.~Isenberg,
  Class.\ Quant.\ Grav.\  {\bf 25}, 195015 (2008).
  [arXiv:0805.1451 [gr-qc]].
  
\bibitem{Friedman:1993ty}
  J.~L.~Friedman, K.~Schleich, D.~M.~Witt,
  Phys.\ Rev.\ Lett.\  {\bf 71}, 1486-1489 (1993).
  [gr-qc/9305017].
  
\bibitem{Ruffini:1971xx}
  R.~Ruffini, J.~Wheeler,
  Physics\ Today\  {\bf 24}, 30 (1967).
  
\bibitem{Bhattacharya:2007ap}
  S.~Bhattacharya, A.~Lahiri,
  Phys.\ Rev.\ Lett.\  {\bf 99}, 201101 (2007).
  [gr-qc/0702006 [GR-QC]].

\bibitem{Dias:2011at}
  O.~J.~C.~Dias, G.~T.~Horowitz, J.~E.~Santos,
  JHEP {\bf 1107}, 115 (2011).
  [arXiv:1105.4167 [hep-th]].
  
\bibitem{Myers:1986un}
  R.~C.~Myers, M.~J.~Perry,
  Annals Phys.\  {\bf 172}, 304 (1986).
  
\bibitem{Hawking:1998kw}
  S.~W.~Hawking, C.~J.~Hunter and M.~Taylor,
  Phys.\ Rev.\  D {\bf 59}, 064005 (1999)
  [arXiv:hep-th/9811056].

\bibitem{Gibbons:2004js}
  G.~W.~Gibbons, H.~Lu, D.~N.~Page and C.~N.~Pope,
  Phys.\ Rev.\ Lett.\  {\bf 93}, 171102 (2004)
  [arXiv:hep-th/0409155].
  
\bibitem{Gibbons:2004uw}
  G.~W.~Gibbons, H.~Lu, D.~N.~Page and C.~N.~Pope,
  J.\ Geom.\ Phys.\  {\bf 53}, 49 (2005)
  [arXiv:hep-th/0404008].

\bibitem{Pena:1997cy}
  I.~Pena, D.~Sudarsky,
  Class.\ Quant.\ Grav.\  {\bf 14}, 3131-3134 (1997).
  
\bibitem{Astefanesei:2003qy}
  D.~Astefanesei, E.~Radu,
  Nucl.\ Phys.\  {\bf B665}, 594-622 (2003).
  [gr-qc/0309131].
  
\bibitem{Hartmann:2010pm}
  B.~Hartmann, B.~Kleihaus, J.~Kunz, M.~List,
  Phys.\ Rev.\  {\bf D82}, 084022 (2010).
  [arXiv:1008.3137 [gr-qc]].
  
\bibitem{Dias:private}
  Private communication with the authors.
  
\bibitem{Wald:1993ki}
  R.~M.~Wald,
  [gr-qc/9305022].
  
\bibitem{Ashtekar:1999jx}
  A.~Ashtekar, S.~Das,
  Class.\ Quant.\ Grav.\  {\bf 17}, L17-L30 (2000).
  [hep-th/9911230].
  
  \bibitem{Das:2000cu}
  S.~Das, R.~B.~Mann,
  JHEP {\bf 0008}, 033 (2000).
  [hep-th/0008028].
  
\bibitem{Gibbons:2004ai}
  G.~W.~Gibbons, M.~J.~Perry, C.~N.~Pope,
  Class.\ Quant.\ Grav.\  {\bf 22}, 1503-1526 (2005).
  [hep-th/0408217].
  


\end{thebibliography}

\end{document}